\documentclass[%
preprint,
 amsmath,amssymb,
 aps,
]{revtex4-2}

\usepackage{float}
\usepackage{xcolor}
\usepackage{graphicx}
\usepackage{dcolumn}
\usepackage{bm}
\usepackage{soul}
\begin{document}

\title{Master Stability Functions of Networks of Izhikevich Neurons}
	
\author{Raul P. Aristides}
 \email{raul.palma@unesp.br}
 \affiliation{S\~ao Paulo State University (UNESP), Instituto de F\'{i}sica Te\'{o}rica, Rua Dr. 
			Bento Teobaldo Ferraz 271, Bloco II, Barra Funda, 01140-070 S\~ao Paulo, Brazil.}
\author{Hilda A. Cerdeira}%
\affiliation{S\~ao Paulo State University (UNESP), Instituto de F\'{i}sica Te\'{o}rica, Rua Dr. 
			Bento Teobaldo Ferraz 271, Bloco II, Barra Funda, 01140-070 S\~ao Paulo, Brazil.}
\affiliation{Epistemic, G\'{o}mez $\&$ G\'{o}mez Ltda. ME, Rua Paulo Franco 520, Vila Leopoldina, 05305-031 S\~ao Paulo, Brazil}%
	
\date{\today}
	
\begin{abstract}
	
Synchronization has attracted the interest of many areas where the systems under study can be described by complex networks. Among such areas is neuroscience, where is hypothesized that synchronization plays a role in many functions and dysfunctions of the brain. We study the linear stability of synchronized states in networks of Izhikevich neurons using Master Stability Functions, and to accomplish that, we exploit the formalism of saltation matrices. Such a tool allows us to calculate the Lyapunov exponents of the Master Stability Function (MSF) properly since the Izhikevich model displays a discontinuity within its spikes. We consider both electrical and chemical couplings, as well as  {global} and  {cluster} synchronized states. The MSFs' calculations are compared with a measure of the synchronization error for simulated networks. We give special attention to the case of electric and chemical coupling, where a riddled basin of attraction makes the synchronized solution more sensitive to perturbations.
\end{abstract}
\maketitle
\section{Introduction}

    Neurons are the building blocks of the brain. These structures can display a rich variety of dynamics depending on their type and the inputs that it receives from other neurons \cite{purves}. Although many features of neurons remain unclear, models are capable of reproducing their observed dynamics. The seminal work of Hodgkin-Huxley (HH) introduced a neuron model based on the giant axon of the squid \cite{hodhux}. Being biophysically accurate, this model became a standard for ODE-based neuron models. Naturally, the collective dynamics of neurons are also of interest, and unfortunately, networks of complex models like the HH can be difficult to deal with both, analytically and computationally. With this in mind, E. M. Izhikevich proposed a neuron model that combines biological features of complex models, like the HH, with computational efficiency \cite{izk1}. 

    Due to these characteristics, the Izhikevich model has been implemented in the study of networks of spiking neurons, including large-scale models \cite{izk2,modolo}. Such models can be used to deepen our understanding of how the interplay between synaptic and neuronal processes produces collective behaviors.  Of great interest is the emergence of rhythms and synchronization of neural activity, as it suggests that the high-dimensional dynamics of neuronal networks can collapse into low-dimensional oscillatory modes \cite{buzs}. Although rhythmic activity and synchronization are often associated with brain pathologies like Parkinson’s disease and epilepsy \cite{jiruska,uhlhaas}, evidence suggests that these two mechanics are involved in information processing, memory formation, and cognition \cite{cohen,axmacher,fries}.

    Since the mathematical description of synchronization is well established in nonlinear dynamics, {it can be applied} to many natural phenomena, including neural activity.
	A central question regarding synchronization theory is whether a synchronized state is stable or not. Among the many contributions to this topic over the last decades, here we focus on the \textit{Master Stability Function} (MSF) formalism \cite{pecora}. The main result of such formalism is a diagonalized variational equation, which allows us to calculate the Lyapunov exponents associated with perturbations transverse to the synchronized manifold. In the original work, it was made possible by assuming that the system can be locally linearized around the synchronized solution, as well as a diagonalizable coupling matrix. Fortunately, this formalism has been extended to a variety of cases including near-identical systems, non-diagonalizable, and multilayer networks \cite{nishik, jsun, della}. Moreover, the stability of partial synchronization, which includes cluster synchronized and chimera states have also been studied in the context of master stability functions \cite{pecora2,sorrentino,schaub,yzhang}.
	
	In this paper, we apply the MSF formalism to study the linear stability of synchronized states in networks of Izhikevich neurons. Both global and {cluster} synchronization patterns are considered. To  {surmount the difficulties that the discontinuities of the Izhikevich model present to} calculate the associated Lyapunov exponents, we use the \textit{Saltation Matrices} formalism. First, we consider the case of global synchronization in networks with electrical coupling, followed by the case with chemical coupling and then with both coupling schemes. For each case, we study how the stability of the synchronized state depends on the coupling strength. We verify that the MSF formalism is effective for probing the threshold for synchronization, except for the last case, where the outcome of the system depends on its initial conditions. This can be the case when the system under study has a \textit{riddled basin} of attraction \cite{alex}. As a result, the system evolves towards a non-synchronized solution even with initial conditions in the neighborhood of the synchronization state \cite{huang}.

\section{MSF applied to Izhikevich model}

The Izhikevich model (IM) is a two-dimensional neuronal model, celebrated for its biophysical accuracy, and its capacity of displaying many spiking patterns without losing computational efficiency. Letting $\mathbf{x} = [x, y]^T$, the model consists of a system of differential equations \cite{izk1}
\begin{equation}
\mathbf{\dot{x}} =  \mathbf{F}(\mathbf{x}) = 
    \begin{bmatrix}
    0.04x^2 + 5x + 140 - y + I \\
    a(bx - y)
    \end{bmatrix}
    \label{izk}
\end{equation}

with the after-spike reset given by 

\begin{equation}
    \text{if}\quad x > 30\text{mV}, \quad\text{then}
    \begin{cases}
      x \rightarrow c \\
      y \rightarrow y + d
    \end{cases}
    \label{reset}
\end{equation}

The variables $x$ and $y$ represent the membrane potential and a membrane recovery variable respectively. {$a$ is associated with the time scale of $y$ while $b$ describes the sensibility of $x$ to subthreshold oscillations of $y$. The parameters $c$ and $d$ are related to the after-spike dynamics of the variables $x$ and $y$.} The parameter $I$ takes into account synaptic currents or injected dc-currents \cite{izk1}. For all calculations, we used $a=0.2$, $b=2$, $c = -56$, $d = -16$ and $I = -99$, {with this choice of parameters, the IM displays chaotic dynamics, as discussed in Nobukawa et al.  \cite{nobukawa}.}  {As in \cite{izk1}, the variables $x$ and $y$, and all the parameters of the model are dimensionless.}

The Izhikevich model can be seen as a hybrid model, a continuous-time evolution process interfaced with a logical process, in this case, the resetting function when $x$ crosses the threshold. These discontinuities interfere with the usual straightforward methods \cite{shimada,benettin,wolf85} to compute the Lyapunov exponents of such a system.

Following Bizarri et al. \cite{biz} we resort to saltation matrices, which allow us to calculate the Lyapunov exponents associated with the master stability functions of networks of Izhikevich neurons. In this framework, these matrices are used as correction factors at the instants of the discontinuities due to the resetting function, yielding the correct Lyapunov exponents.

Between spikes the dynamics of the Izhikevich is smooth, and for a small perturbation $\delta\mathbf{x}$ we have
\begin{equation}
    \delta\dot{\mathbf{x}} = {D\mathbf{F}\big|_{\mathbf{x}}} \delta\mathbf{x} \ ,
\end{equation}
where {$D\mathbf{F}\big|_{\mathbf{x}}$ is the Jacobian matrix applied to $F$ evaluated at $\mathbf{x}$}. If a spike occurs, the after-spike reset \eqref{reset} induces a discontinuity in the flow such that {$D\mathbf{F}\big|_{\mathbf{x}^-} \neq D\mathbf{F}\big|_{\mathbf{x}^+}$}, where $-$ and $+$ indicates the before and after the reset. Then, the perturbation $\delta\mathbf{x}^+$ after the reset event is given by
\begin{equation}
    \delta\mathbf{x}^+ = S ({D\mathbf{F}\big|_{\mathbf{x^-}}}\delta\mathbf{x}^-) \ ,
\end{equation}
which allows us to write
\begin{equation}
    \dot{\delta\mathbf{x}^+} = {D\mathbf{F}\big|_{\mathbf{x}^+}}\delta\mathbf{x}^+ \ ,
\end{equation}
where $S$ is the \textit{saltation matrix}, which for Izhikevich neurons is given by
\begin{equation}
    S = 
 \begin{bmatrix}
    \dfrac{\strut \dot{x}^+}{\strut \dot{x}^-} & \strut 0 \\
\dfrac{\strut \dot{y}^+ - \dot{y}^- }{\strut \dot{x}^-} & \strut 1 
\end{bmatrix} \ .
\label{salta_izk}
\end{equation}
The complete derivation of the procedure can be found at \cite{bernardo,biz}.

{Throughout this study, we numerically solve Eq.~\eqref{izk} with a backward differentiation formula, which yields a variable integration step, with a maximal step of $dt = 0.001$. The initial conditions for the simulations, unless stated differently, were picked from a normal distribution centered at a fixed point of Eq.~\eqref{izk} in the case of no coupling, $(-56.25,-112.5)$, with a standard deviation equal to $1$.}
\subsection{Electrical coupling}

First, we study the effect of electrical coupling, which represents electrical synapses between nodes \cite{perkel}. 
Consider a network with $N$ nodes, where the isolated dynamics of each node is given by Eq. \eqref{izk}. An adjacency matrix $A$ indicates whether node $i$ is connected diffusively to node $j$.  {If the nodes are coupled through their first variable,} the dynamics of each node is modified by
\begin{equation}
    W_i = - g_e \sum_{j=1}^N A_{ij} (x^j - x^i)
\end{equation}
where $g_e$ is the coupling conductance (strength) and $A_{ij}$ is an element of the adjacency matrix. Note that in this case, the coupling can be written as a function of the Laplacian matrix of the network, given by $L = {D} - A$ where $D$ is the degree matrix defined by $D_{ii} = \sum_{j} A_{ij}$, then $W_i = g_e\sum_{j=1}^N L_{ij}x^j$. If we define $\mathbf{x} = [\mathbf{x}^1,...,\mathbf{x}^N]^T$, where $\mathbf{x}^i = [x^i,y^i]^T$, and $\mathbf{F}(\mathbf{x}) = [\mathbf{F}(\mathbf{x}^1),...,\mathbf{F}(\mathbf{x}^N)]^T$, we can write the dynamical system of the network as
\begin{equation}
    \mathbf{\dot{x}} =  \mathbf{F}(\mathbf{x}) - g_e (L \otimes \mathbf{G})\mathbf{x}
    \label{izkelec}
\end{equation} 
where $\otimes$ is the Kronecker product and
\begin{equation}
    \mathbf{G} =
    \begin{bmatrix}
    1 & 0  \\
    0 & 0 
    \end{bmatrix}
\end{equation}
is the matrix encoding the coupling scheme.
The synchronization manifold $\mathbf{x}^s$ is defined by the $N-1$ constraints  $\mathbf{x}^1 = \mathbf{x}^2 = ... = \mathbf{x}^N$, and since the Laplacian matrix is a zero row-sum matrix, $(L \otimes \mathbf{G})\mathbf{x^s}$ = 0, we have $\mathbf{\dot{x}^s} = \mathbf{F}(\mathbf{x}^s)$.
The linear stability of the synchronized solution can be investigated with the MSF of Eq. \eqref{izkelec}
\begin{equation}
    \dot{\eta}_k = [{D\mathbf{F}\big|_{\mathbf{x}^s}} - g_e \gamma_k {D\mathbf{G}\big|_{\mathbf{x}^s}}] \eta_k \ ,
    \label{izmsfdif}
\end{equation}
where $\gamma_k$ is the $k$-th eigenvalue of $L$. Thus, the analysis is decomposed in eigenmodes, one being parallel to the synchronization manifold $\gamma_0$, the others transverse to it. We are interested in the Lyapunov exponents of transverse modes, which tell us whether or not perturbations to the synchronized solution will decay. Equation \eqref{izmsfdif} was first derived by Pecora and Carroll \cite{pecora} for undirected networks ($A = A^T$) but since then it has been extended to a broad range of networks and synchronization patterns \cite{nishik, jsun, della,pecora2,sorrentino,schaub,yzhang}. {We highlight that at the event of a spike, the synchronized solution $\mathbf{x}^s$ encounters a discontinuity, and therefore at this point, we need to apply the saltation matrix Eq. \eqref{salta_izk}}.
\begin{figure}[H]
    \centering
    \includegraphics[width=0.5\textwidth]{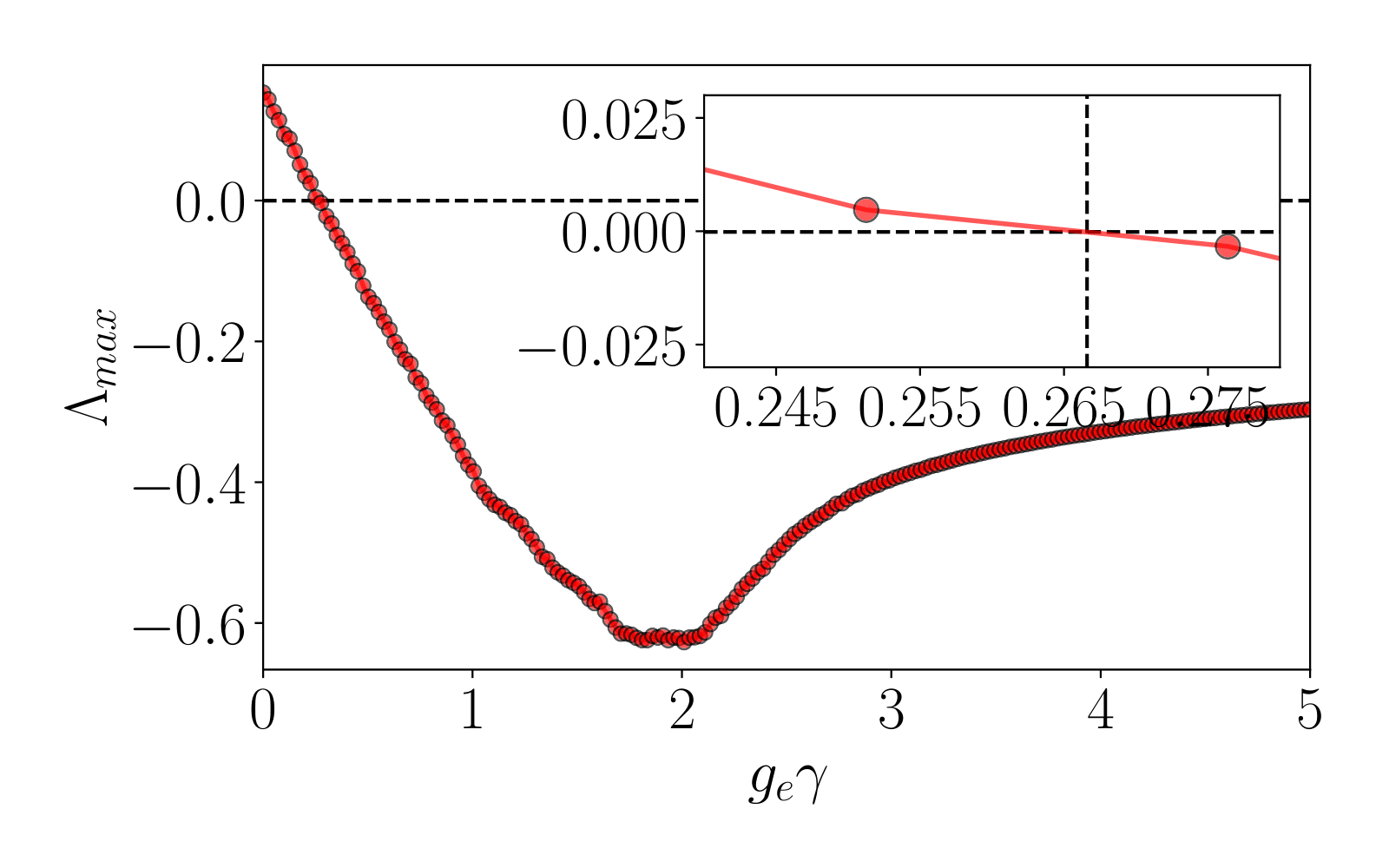}
    \caption{ {Maximum Lyapunov exponent transversal to the synchronization manifold for electrically coupled Izhikevich neurons as a function of  $g_e\gamma$.}}
    \label{msf1}
\end{figure}

\begin{figure}[H]
    \centering
    \includegraphics[width=0.2\textwidth]{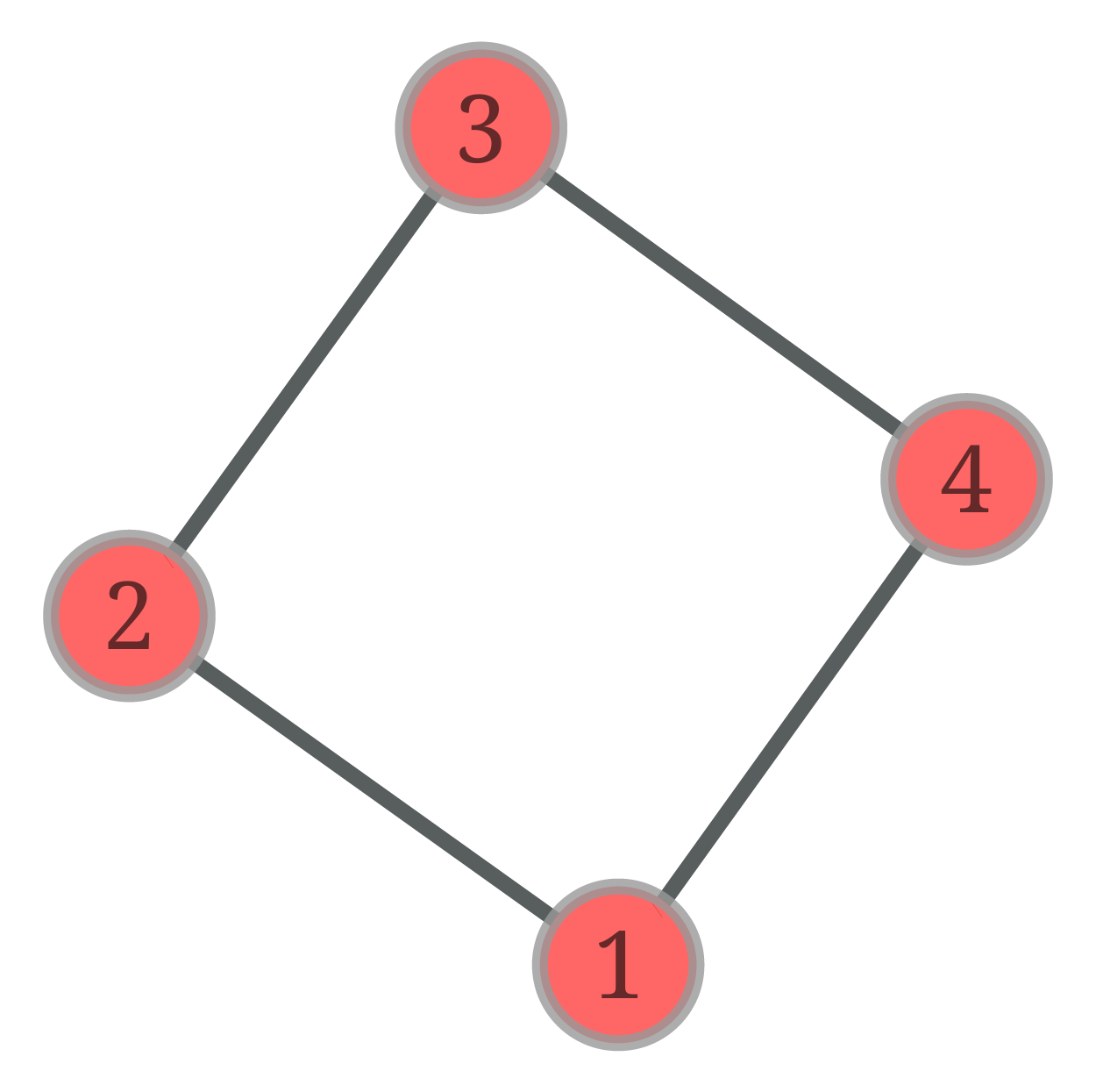}
    \caption{ {Undirected network with $N = 4$ nodes, which is used to exemplify the stability of complete synchronization for identical Izhikevich neurons.}}
    \label{net_n4}
\end{figure}

The maximum Lyapunov exponent (MLE) of Eq. \eqref{izmsfdif} is presented in Fig. \ref{msf1}. The expanded inset shows that $\Lambda_{max}$ becomes negative for $g_e\gamma \gtrsim 0.267$. {To check our results, we simulate a network of $N=4$ nodes, as depicted in Fig. \ref{net_n4}. In such a case, the Laplacian matrix is}
\begin{equation}
    L =
    \begin{bmatrix}
    2 & -1 & 0 &  -1  \\
    -1 & 2 & -1 &  0  \\
    0 & -1 & 2 &  -1  \\
    -1 & 0 & -1 &  2  \\
    \end{bmatrix}
\end{equation}
with eigenvalues  $0, 2, 2$ and $4$. {The smallest eigenvalue $\gamma_1 = 0$} {is associated with the synchronized manifold, and the other two are transverse to it}.  {The degeneracy in the eigenvalues affects the stability of the clusters associated with them, meaning that the clusters associated with $\gamma_i= 2$ will become linearly stable for the same value of the coupling strength.} Putting together that the MLE is negative for $g_e \gamma > 0.2670$, and that $2$ is the {smallest} eigenvalue associated with the transverse mode, we conclude that for $g_e > 0.133$ the synchronized solution is stable.

\begin{figure}[H]
    \centering
    \includegraphics[width=0.5\textwidth]{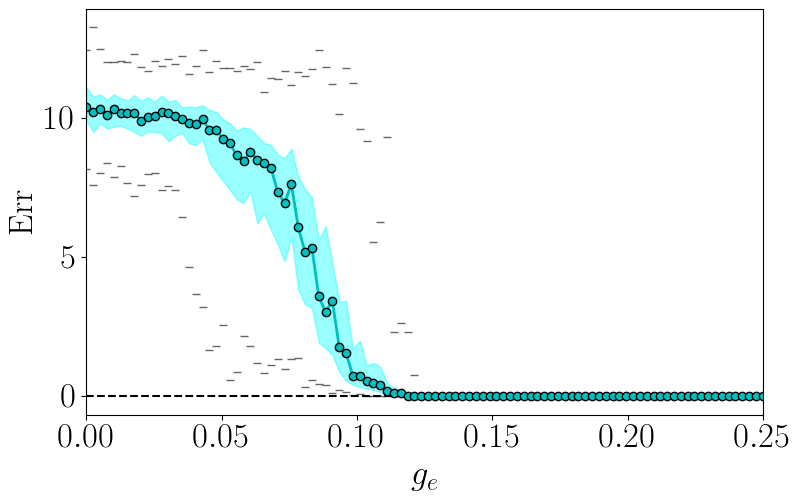}
    \caption{ {Synchronization error for a network of $N = 4$ Izhikevich neurons  with electrical coupling. It goes to zero at $g_e \approx 0.133$, in conformity with the MSF. The solid line represents the average of $100$ simulations, the shaded area represents the first and third quartiles, and the grey dashes represent the maximum and minimum values.}}
    \label{err1}
\end{figure}

Figure \ref{err1} depicts the synchronization error, which is computed as {$\sum_{j}^N |\mathbf{\bar{x}} - \mathbf{x}^j|$ where $\mathbf{\bar{x}}$ is the average state for the nodes in the networks. We notice a} good agreement with the result given by the MSF {analysis}.

\subsection{Chemical coupling}

We now study the synchronization of an Izhikevich network with chemical synaptic coupling between nodes. The chemical synapses that node $i$ receives from every other node $j$ are modeled by the sigmoidal function $B_i$ \cite{somers}
\begin{equation}
    B_i = g_c (x^i - v_s) \sum_{j = 1}^N A_{ij} \zeta(x^j),
    \label{chem}
\end{equation}

where $v_s$ is the reversal potential,  $\zeta(x) = [1 + \exp(-\epsilon(x-\theta))]^{-1}$, $\epsilon$ defines the slope of the sigmoidal function and $\theta$ is the synaptic firing threshold. Througout this study we set {$v_s = 0$}, $\epsilon = 7$ and $\theta = 0$. This set of parameters allows the chemical synapses to be both excitatory and inhibitory, since $v_s$ lies inside the range of oscillation of the action potential $x_i$. This is the case for some classes of neurons, {as discussed in} \cite{root,pelkey}.

Note that this coupling term cannot be written as a function of the Laplacian matrix, instead we can write it in terms of the adjacency matrix.  To obtain an equation analogous to \eqref{izkelec}, we introduce a few objects, which are defined in the Appendix \ref{Appx}, together with the complete derivation of the equation for the dynamics of the network with chemical coupling, which is given by: 

\begin{equation}
    \mathbf{\dot{x}} = \mathbf{F}(\mathbf{x}) - g_c \mathbf{T}(\mathbf{x})[(A \otimes \mathbf{C})(\mathbf{x})] \ .
\end{equation}

The synchronized solution, $\mathbf{x}^1 = \mathbf{x}^2 = ... = \mathbf{x}^N$, derived in the Appendix \ref{Appx} from Eq. \eqref{appchem0} to \eqref{appchem1}, takes the form \begin{equation}
    \mathbf{\dot{x}}^s = \mathbf{F}(\mathbf{x}^s) - g_c k_n \mathbf{T}(\mathbf{x}^s)\mathbf{C}(\mathbf{x}^s) \ ,
\end{equation}
and it exists only if the number of links $k_n$ is the same for all nodes \cite{checco}. Following the Pecora-Carroll analysis, we obtain the Master Stability Function for this case, and after we diagonalize $A$ we have
\begin{align}
    \dot{\eta} = \{[\mathbf{I}_N \otimes ({D\mathbf{F}\big|_{\mathbf{x}^s}} - g_ck_n{D\mathbf{H}\big|_{\mathbf{x}^s}}\mathbf{K}(\mathbf{x}^s))]   - g_c \Gamma^A \otimes \mathbf{T}(\mathbf{x}^s){D\mathbf{C}\big|_{\mathbf{x}^s}} \} \eta \ .
    \label{msfizk2}
\end{align} 
The complete derivation of this equation, which is analogous to Eq. \eqref{izmsfdif}, is given in Appendix \ref{Appx} and it was first derived by Checco et al. \cite{checco}.
As an example, we consider {the} network of $N = 4$ nodes, {depicted in Fig. \ref{net_n4}}, with an adjacency matrix given by
\begin{equation}
A = 
    \begin{bmatrix}
       0 & 1 & 0 & 1 \\
       1 & 0 & 1 & 0 \\
       0 & 1 & 0 & 1 \\
       1 & 0 & 1 & 0
    \end{bmatrix}
    \label{izk_N4}
\end{equation}

\begin{figure}[H]
    \centering
    \includegraphics[width=0.5\textwidth]{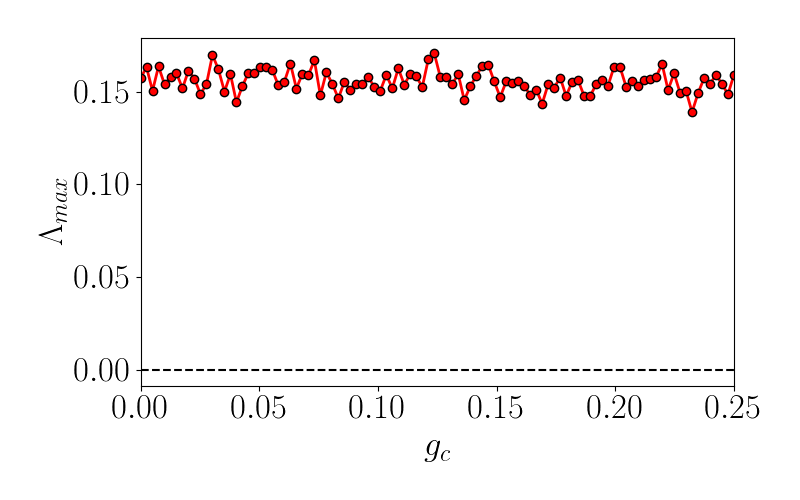}
    \caption{ {Maximum Lyapunov exponent transversal to the synchronization manifold ($\gamma_4 = -2$) for $N=4$ for chemically coupled Izhikevich neurons.}}
    \label{msf_izk2}
\end{figure}

\begin{figure}[H]
    \centering
    \includegraphics[width=0.5\textwidth]{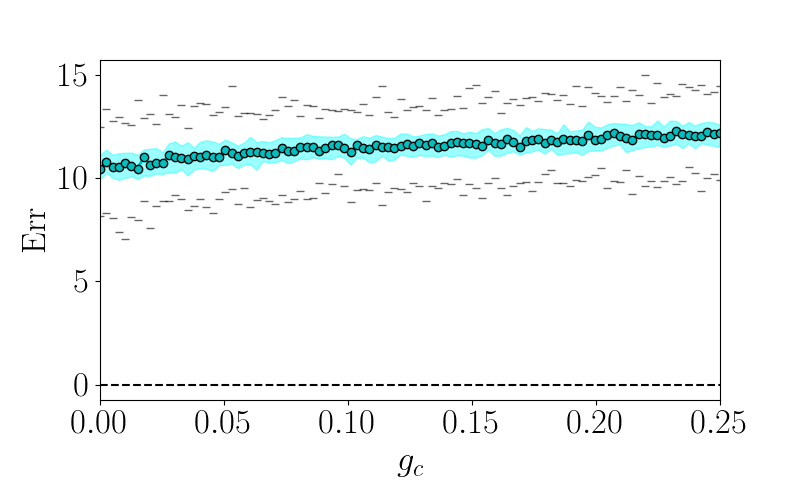}
    \caption{ {Synchronization error for a network of $N = 4$ Izhikevich neurons  with chemical coupling. As expected from the MSF analysis, the system does not synchronize. The solid line represents the
average of 100 simulations, the shaded area represents the first and third quartiles, and the grey dashes
represent the maximum and minimum values.}}
    \label{err_izk2}
\end{figure}

The adjacency matrix Eq. \eqref{izk_N4} has an eigenvalue $\gamma_1 = 2$  associated with the synchronized solution,  {since its associated eigenvector is proportional to the unit vector. Hence} we need to calculate the MLE for the remaining eigenvalues $\{0,0,-2\}$.  {Unlike Eq. \eqref{izmsfdif}, in Eq. \eqref{msfizk2} we cannot study it as a function of $g_c\Gamma^A$, which would give us an insight into all the possible combinations of networks and coupling strengths. This is so because of the presence of the term $g_ck_n$, which is the row-sum of the matrix $A$.} Figure \ref{msf_izk2} shows the MLE of Eq. \eqref{msfizk2}  {for $\gamma_4 = -2$,} which is positive for the range studied. Our calculation for the synchronization error in Fig. (\ref{err_izk2}) shows good agreement with the MLE {, i.e., we verified that the network does not synchronize with chemical coupling}. 

\subsection{Chemical and electrical coupling}

Finally, we study the effect of both electrical and chemical coupling on synchronization. In this case, the evolution of the network is given by
\begin{equation}
    \mathbf{\dot{x}} = \mathbf{F}(\mathbf{x}) - g_e L\otimes\mathbf{G}(\mathbf{x}) - g_c \mathbf{T}(\mathbf{x})(A \otimes \mathbf{C})\mathbf{x}  \ . 
\end{equation} 
For the sake of simplicity, we consider that both interactions depend on the same diagonalizable adjacency matrix, i.e., $L = D - A$, and that every node has the same number of neighbors $k_n$ {, meaning that $A$ is a regular matrix}. Combining the results of previous sections, we write the variational equation for the perturbations to the synchronized state as
\begin{align}
    \delta\mathbf{\dot{x}} = \{[\mathbf{I}_N \otimes ({D\mathbf{F}\big|_{\mathbf{x}^s}} - g_ck_n{D\mathbf{H}\big|_{\mathbf{x}^s}} \mathbf{K}(\mathbf{x}^s))]  - g_e L\otimes {D\mathbf{G}\big|_{\mathbf{x}^s}}  - g_c A \otimes \mathbf{T}(\mathbf{x}^s){D\mathbf{C}\big|_{\mathbf{x}^s}} \} \delta\mathbf{x}  \ .
\end{align}
Since $L$ and  $A$ commute, we can diagonalize both simultaneously, yielding
\begin{align}
\dot{\eta} = \{[\mathbf{I}_N \otimes ({D\mathbf{F}\big|_{\mathbf{x}^s}} - g_ck_n{D\mathbf{H}\big|_{\mathbf{x}^s}} \mathbf{K}(\mathbf{x}^s))]  - [g_e \Gamma^L\otimes {D\mathbf{G}\big|_{\mathbf{x}^s}} - g_c \Gamma^A \otimes \mathbf{T}(\mathbf{x}^s){D\mathbf{C}\big|_{\mathbf{x}^s}}] \} \eta  \ .
   \label{msf_izk3}
\end{align} 

where $\Gamma^L$ and $\Gamma^A$ are diagonal matrices with eigenvalues of $L$ and $A$ respectively as entries.

\begin{figure}[H]
    \centering
    \includegraphics[width=0.5\textwidth]{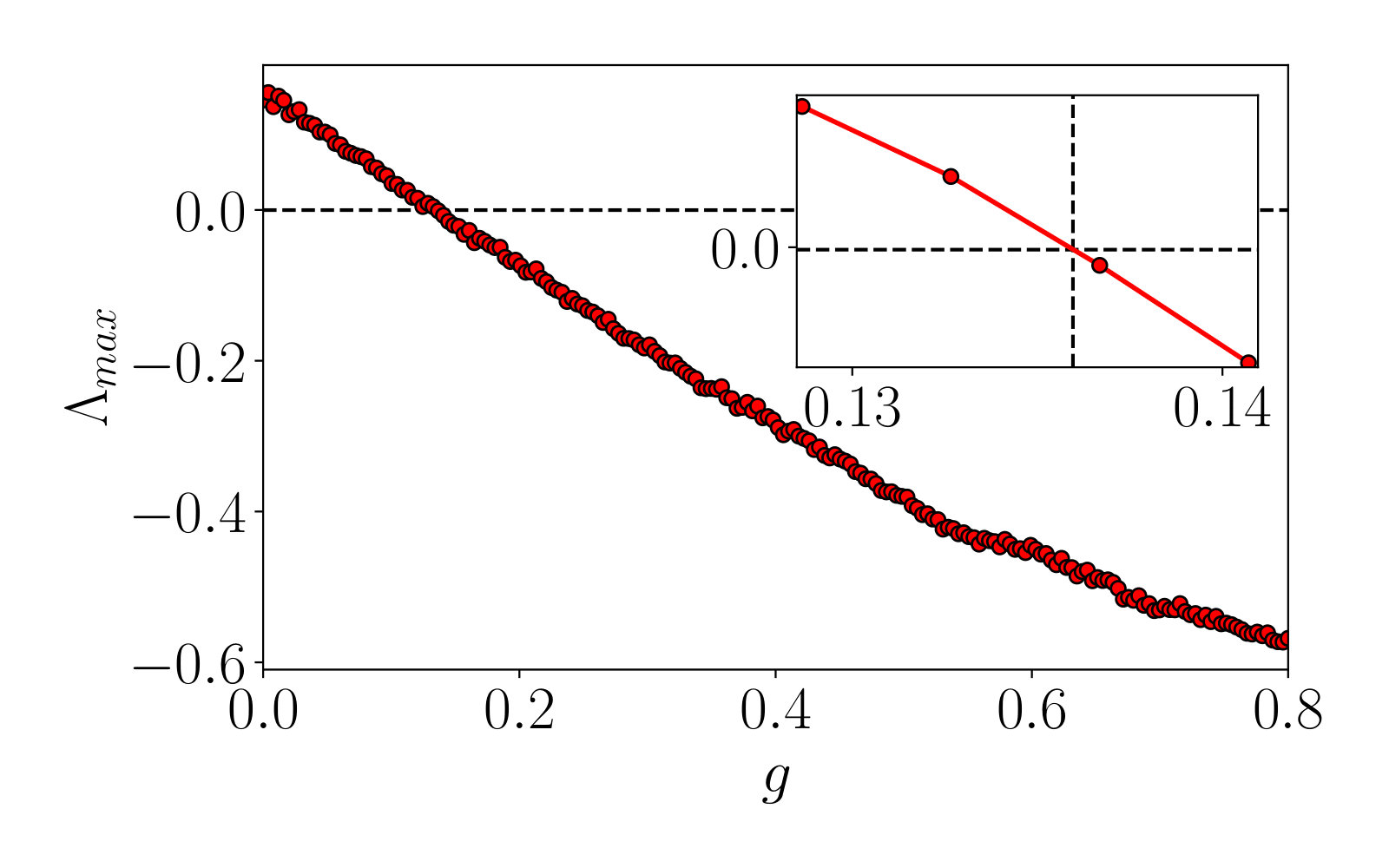}
    \caption{ {Maximum Lyapunov exponent transversal to the synchronization manifold for $N=4$ Izhikevich neurons with electrical and chemical coupling.}}
    \label{msf3}
\end{figure}

\begin{figure}[H]
    \centering
    \includegraphics[width=0.5\textwidth]{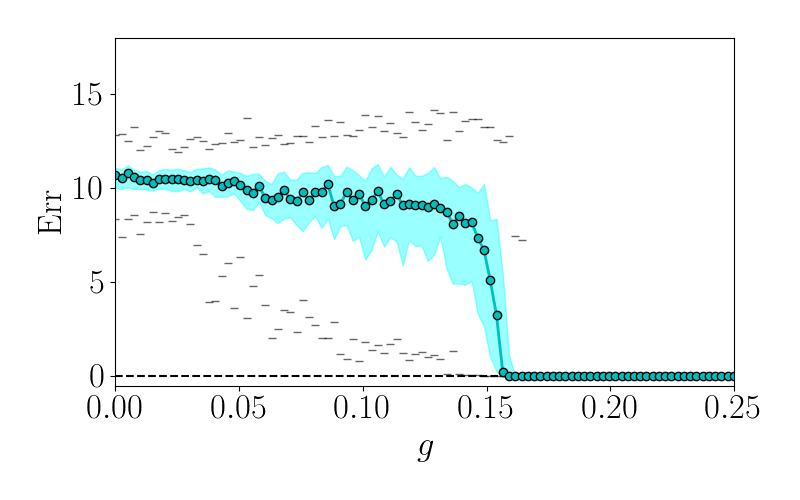}
    \caption{ {Synchronization error for $N=4$ with electrical and chemical coupling. The solid line represents the
average of 200 simulations, the shaded area represents the first and third quartiles and the grey dashes
represent the maximum and minimum values.}}
    \label{err3}
\end{figure}

Once again, we take a network of $N=4$ neurons, with electrical and chemical coupling given by $g = g_e = g_c$, {and represented by the} adjacency matrix  \eqref{izk_N4}.
Thus, the eigenvalues associated with transverse modes are $\{\gamma^A\} = \{0,0,\-2\} $ from the adjacency matrix and $\{\gamma^L\} = \{2,2,4\}$, from the Laplacian matrix.
We calculate the MLE of Eq. \eqref{msf_izk3} as a function of $g$, and as shown in Fig. \ref{msf3}, the global synchronization pattern becomes  {linearly} stable for $g \approx 0.13$. To confirm this result we calculate the synchronization error over $200$ simulations, which is shown in Fig. \ref{err3}, and find that it does not go to zero when the MLE becomes negative.

We notice that the synchronization error only vanishes around  {$g \approx 0.16$}, and for values below that, the system can either synchronize or not. To better understand these results, we calculate the synchronization error for different initial conditions, for {$g = 0.155$}. {While the  $y$-variables are set to $-101.5$, we partitioned the network into two distinct clusters according to their $x$-variables, namely  {$x_1 = x_3$ and $x_2 = x_4$}. The range of $x$-variables spans from $1$ to $-1$, resulting in initial conditions  {$x$-variables} that are either orthogonal, represented as $\pm[1, -1, 1, -1]$, or parallel to the synchronized manifold, denoted as $\pm[1, 1, 1, 1]$. The value $y = -101.5$ corresponds to an approximate value for $y$ in the range of $x$ we selected}.

\begin{figure}[H]
\includegraphics[width=.48\linewidth]{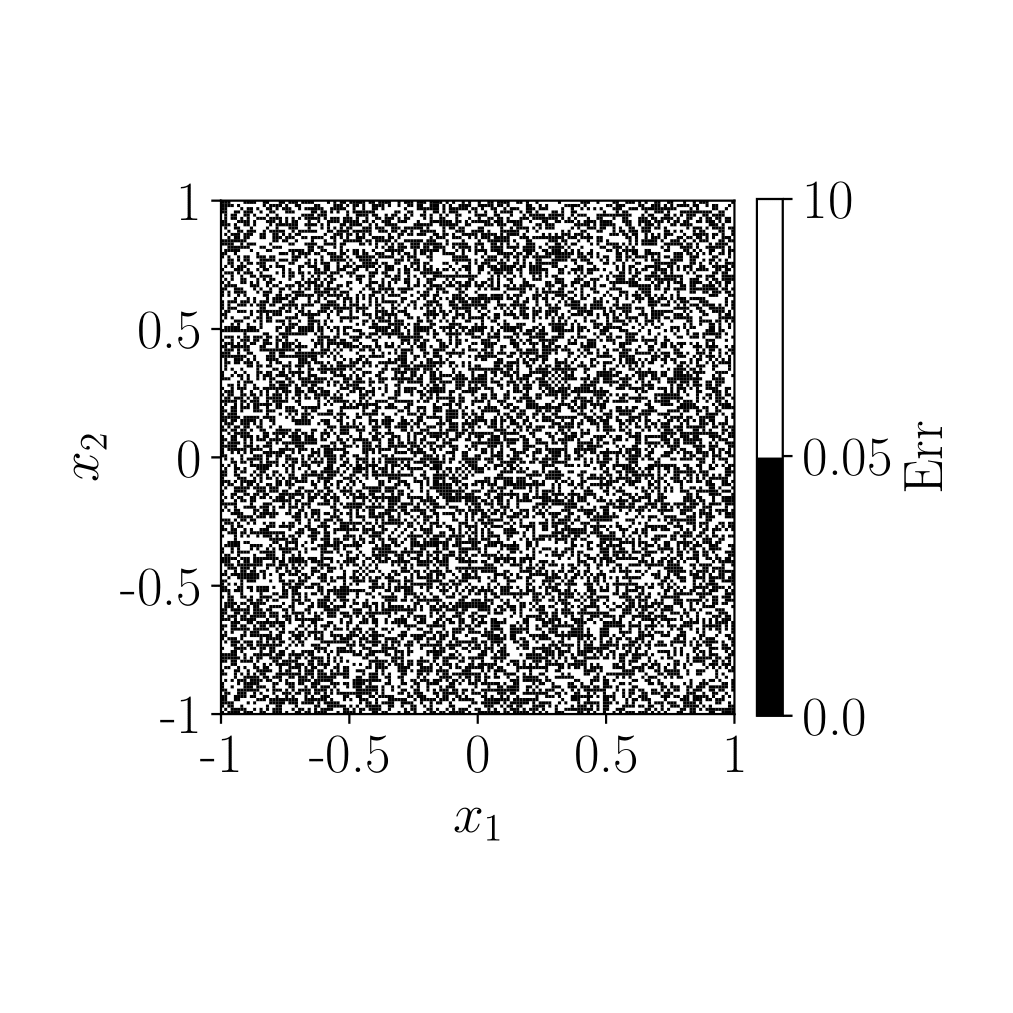}\hfill
\includegraphics[width=.48\linewidth]{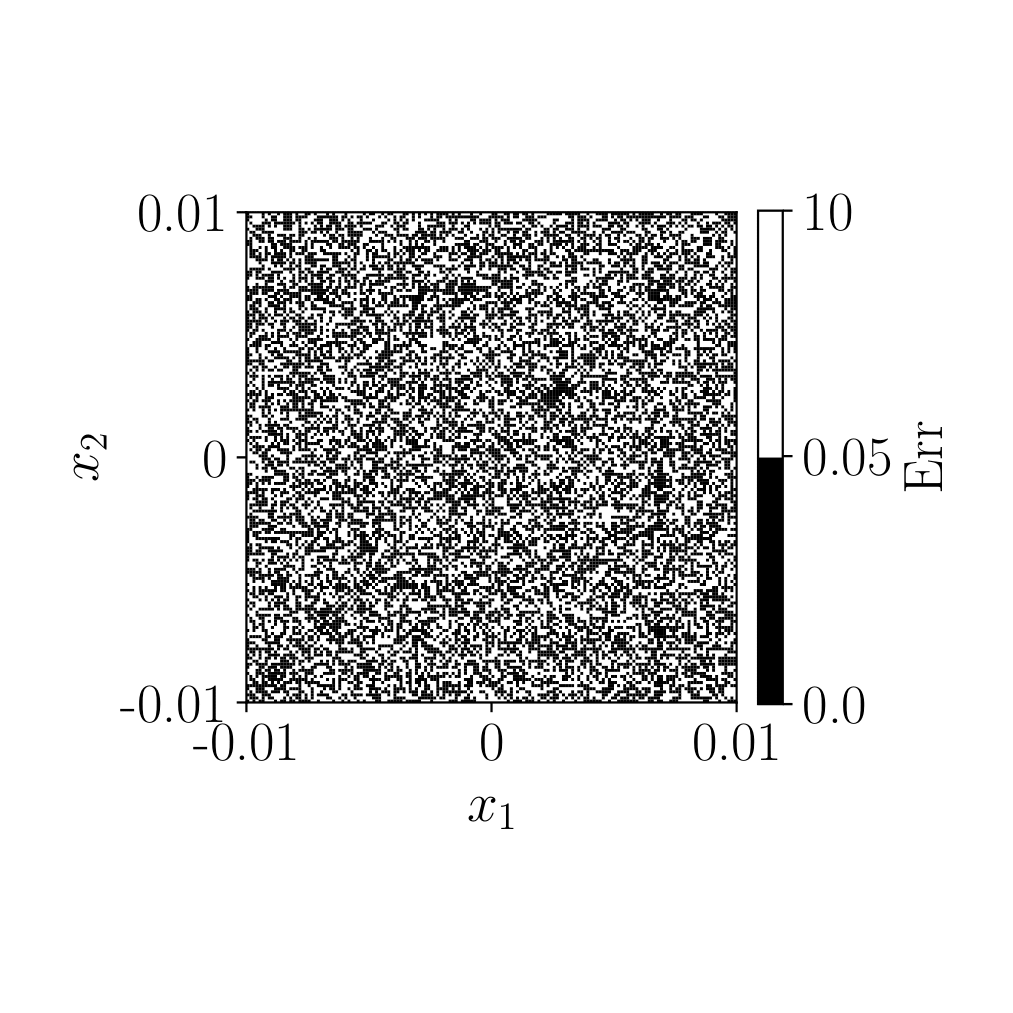}
    \caption{ {Riddled basin of attraction in $(x_1,x_2)$ plane for $g = 0.155$, calculated with $10.24\times10^4$ initial conditions in (a) $x_1, x_2 \in [1,-1]$ and (b) $x_1, x_2 \in [0.01,-0.01]$ {while $x_3 = x_1$, $x_4 = x_2$ and $y$-variables are set to $y_{i}= -101.5$}. Solutions that reach a final state with a synchronization error Err $ > 0.05$ are marked as white dots; otherwise, they are marked as black.}}
    \label{riddbasin0}
\end{figure}

As we see in Fig. \ref{riddbasin0} (a) for $g = 0.155$, depending on the initial conditions, the system can end up in different solutions, a characteristic of multistable systems \cite{ulrk}. Moreover, the different solutions are intertwined in a complex way. This type of basin of attraction is often called a riddled basin, and can be the cause for the divergence between the MLE and the synchronization error results \cite{huang}. {Figure \ref{riddbasin0} (b) shows an inset of Fig. \ref{riddbasin0} (a), where we see that synchronized solutions are intertwined with non-synchronized in a similar {way} as in Fig. \ref{riddbasin0} (a) even though the domain is reduced.}

{Besides that, it is usually assumed that the presence of riddled basins of attraction is related to the MLE approaching zero from below, as discussed in \cite{heagy,ott94}. A similar result, where the system fails to exhibit synchronization when the MLE approaches zeros from above was reported by \cite{huang} but the presence of the riddled basin was not investigated.} 

{To better understand the basin of attraction, we calculate the uncertainty exponent $\alpha$, which as defined in \cite{mcdonald_fss, grebogi_fss}, measures how small perturbations to initial conditions affect the final state of the system. To do so, we pick $M=1000$ pairs of initial conditions with $g = 0.155$, iterating each to its final state. Each pair is separated by a distance $d$, i.e., $\mathbf{x} - \mathbf{x}' = d$. We then compare the final states of each initial condition to verify if there are pairs that end up in different states and therefore are \textit{uncertain}. The fraction of the phase space that is uncertain obeys
\begin{equation}
    f(\varepsilon) = \varepsilon^\alpha
    \label{uncertain}
\end{equation}
with $\alpha < 1$. We record that the uncertainty exponent is related to the dimension of the phase space $D$ as $\alpha = D- \delta$, where $\delta$ is the (capacity) dimension of the basin boundary \cite{grebogi_fss}.}

{We found an uncertainty exponent equal to $\alpha \approx 0.005$. As a result, even for extraordinarily small perturbations to the synchronized state, a fraction of them can grow and desynchronize the system. Moreover, using the $\alpha = D - \delta$, we find that the dimension of the basin boundary, $\delta \approx 7.995$, is close to the dimension of the state space. In conclusion, the obtained uncertainty exponent of $\alpha \approx 0.005$ suggests the presence of a riddled basin of attraction \cite{ujjwal_fss,ulrk}.}

\begin{figure}[H]
    \centering
    \includegraphics[width=0.5\textwidth]{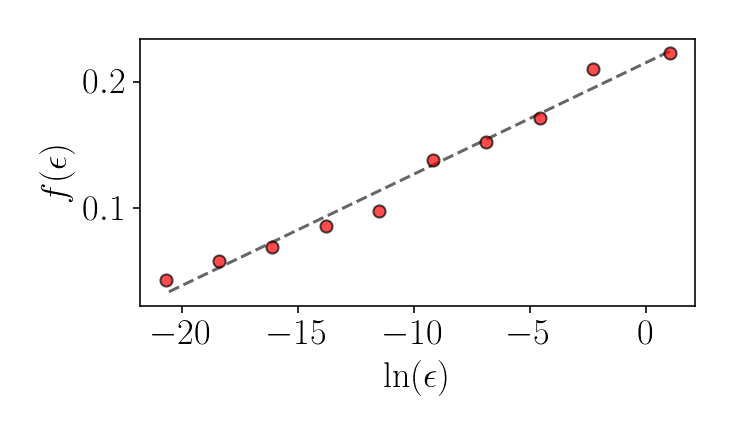}
    \caption{ {The fraction of pairs of initial conditions that converge to different asymptotic solutions, $f$, as a function of the distance between initial condition  $\varepsilon$ - Eq. \eqref{uncertain}. For $g = 0.1555$, the slope of the line that best fits the points yields an uncertainty exponent $\alpha = 0.005$.}}
    \label{}
\end{figure}

\section{Izhikevich model: Cluster Synchronization}

\subsection{Electrical coupling}

As mentioned in the first section, the electrical coupling can be written in terms of the  Laplacian matrix. In this case, the variational equation concerning the linear stability around the cluster synchronized state (CSS) is given by \cite{pecora2}
\begin{equation}
    \delta\dot{\mathbf{x}} = [\sum_{m=1}^M (E^m \otimes {D\mathbf{F}\big|_{\mathbf{s}^m}} - g_e(LE^m \otimes {D\mathbf{G}\big|_{\mathbf{s}^m}}))]\delta\mathbf{x}
    \label{partvar}
\end{equation}
where $M$ is the number of clusters and $E^m$ is an $N\times N$ diagonal indicator matrix for each cluster such that $E^m_{ii} = 1$ if $i$ belongs to cluster $m$, and it is equal to zero otherwise. If we consider a CSS with $M$ clusters, the dynamical evolution of such state is given by \cite{sorrentino,schaub}
\begin{align}
\dot{\mathbf{x}} &= (Z \otimes \mathbf{I}_q)\dot{\mathbf{s}}, \quad\text{where} \\
\dot{\mathbf{s}} &= \mathbf{F}(\mathbf{s}) - g_e(L^{\pi} \otimes \mathbf{I}_q)\mathbf{G}(\mathbf{s})
\label{clsync}
\end{align}
and $Z$ is the $N\times M$ indicator matrix which encodes the cluster state configuration $Z: Z_{ij}=1$ if node $i$ is part of cluster $C_j$ and $Z_{ij} = 0$ otherwise. $\mathbf{I}_q$ is an $q \times q$ identity matrix, where $q$ is the dimension of the isolated dynamical system, $q = 2$ in our case. The  state of each cluster is encoded in the $M\times q$-dimensional vector $\mathbf{s}$. If the quotient network Laplacian $L^\pi$ satisfies the condition \cite{cardoso}
\begin{equation}
    LZ = ZL^\pi \ ,
    \label{sylvstr}
\end{equation}
where $L^{\pi} = (Z^T)^{-1}Z^TLZ = Z^{+}LZ$ and $Z^{+}$ is the left Moore-Penrose of  pseudoinverse $Z$ \cite{moore,penrose}, then the partition encoded by $Z$ is said to be an Equitable External Partition (EEP) \cite{cardoso}. In this case, the network is divided into $M$ clusters in such way that the number of connections from nodes in a cluster $C_i$ to nodes in a cluster $C_j$ depends only on $i,j$ with $i \neq j$. The quotient network dynamics is a coarse-grained version of the original network, where each cluster becomes a node and the weights between these nodes are the out-degrees between clusters in the original network \cite{oclery,schaub}.

\begin{figure}[h!]
    \includegraphics[width=0.25\textwidth]{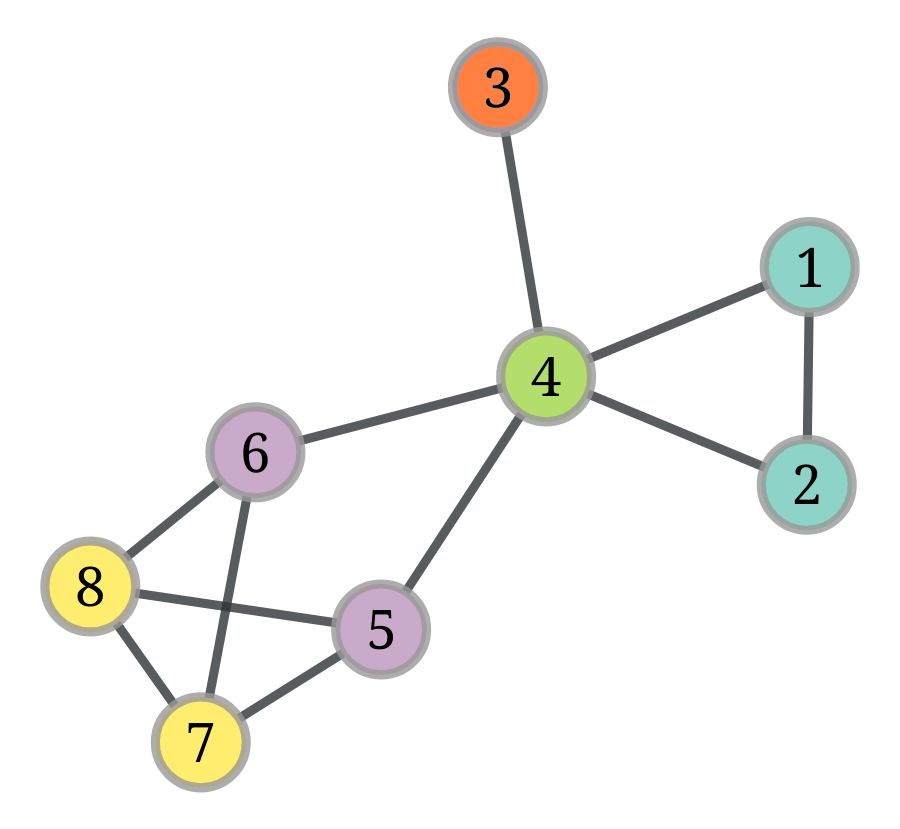}
    \caption{ {Undirected network with N = 8 nodes, which is used to exemplify the stability of cluster synchronization, where the clusters are indicated with colors.}}
    \label{oclery}
\end{figure}

Again, we start our example with a network of Izhikevich neurons coupled through electrical synapses. Consider a network with the following Laplacian matrix
\begin{equation}
    L = 
    \begin{bmatrix}
    2 &  \-1 &   0 &  \-1 &   0 &   0 &   0 &   0  \\
    \-1 &   2 &   0 &  \-1 &   0 &   0 &   0 &   0  \\
     0 &   0 &   1 &  \-1 &   0 &   0 &   0 &   0  \\
    \-1 &  \-1 &  \-1 &   5 &  \-1 &  \-1 &   0 &   0  \\
     0 &   0 &   0 &  \-1 &   3 &   0 &  \-1 &  \-1  \\
     0 &   0 &   0 &  \-1 &   0 &   3 &  \-1 &  \-1  \\
     0 &   0 &   0 &   0 &  \-1 &  \-1 &   3 &  \-1  \\
     0 &   0 &   0 &   0 &  \-1 &  \-1 &  \-1 &   3
    \end{bmatrix} \  
\end{equation}
and a CSS with indicator matrix $Z$ given by  {5 clusters (3 with 2 units, and 2 with 1 unit each},
\begin{equation}
    Z = 
    \begin{bmatrix}
    1 & 0 & 0 & 0 & 0\\
    1 & 0 & 0 & 0 & 0\\
    0 & 1 & 0 & 0 & 0\\
    0 & 0 & 1 & 0 & 0\\
    0 & 0 & 0 & 1 & 0\\
    0 & 0 & 0 & 1 & 0\\
    0 & 0 & 0 & 0 & 1\\
    0 & 0 & 0 & 0 & 1\\
    \end{bmatrix} \ .
    \label{z1}
\end{equation}
One can verify that this choice of $L$ and $Z$ satisfies \eqref{sylvstr} and therefore this cluster configuration is an EEP. A representation of such network is given in Fig. \ref{oclery}, where nodes of the same color represent a cluster. Putting this together, Eq. (\ref{clsync}) reads

\begin{equation}
    [\dot{\mathbf{x}}]^T = [\dot{\mathbf{s}}^1,\dot{\mathbf{s}}^1,\dot{\mathbf{s}}^3,\dot{\mathbf{s}}^4,\dot{\mathbf{s}}^5,\dot{\mathbf{s}}^5,\dot{\mathbf{s}}^7,\dot{\mathbf{s}}^7]^T \ .
\end{equation}

After block-diagonalizing the variational equation \eqref{partvar}, we investigated the linear stability of the CSS with the following equation
\begin{equation}
     \mathbf{\dot{\eta}} = \Big[ \sum_{m} Q^{m} \otimes {D\mathbf{F}\big|_{\mathbf{s}^m}} - g_e \Gamma^L Q^m\otimes {D\mathbf{G}\big|_{\mathbf{s}^m}} \Big] \mathbf{\eta} \ .
     \label{msfelecpartial}
\end{equation}
Where $Q^m = V^{-1}E^m V$ is a block diagonal matrix and $\Gamma^L = V^{-1}LV$ is a diagonal matrix filled with eigenvalues of $L$. The orthogonal matrix that block-diagonalizes both ${E^m}$ and $L$ can be found using group theory \cite{sorrentino}, External Equitable Partitions (EEP's) \cite{schaub} and Simultaneous Block Diagonalization (SBD) \cite{yzhang} approaches,  resulting in the decoupling between modes within the cluster synchronized manifold and transverse to it. 

\begin{figure}[H]
    \centering
    \includegraphics[width=0.5\textwidth]{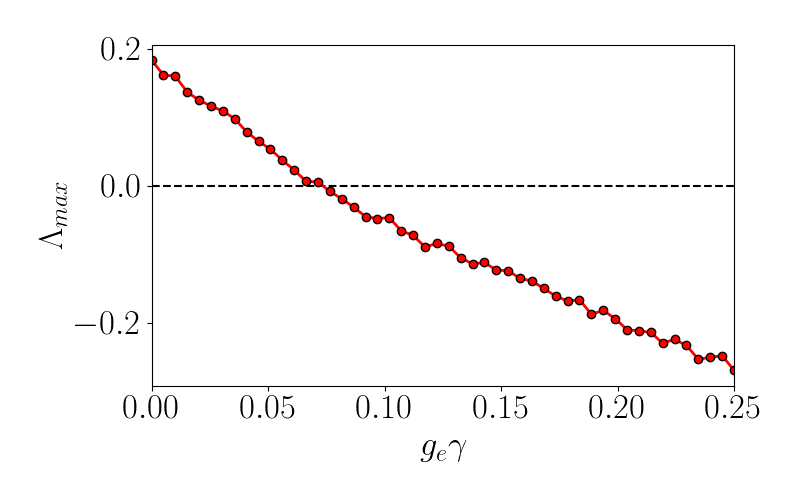}
    \caption{ {Maximum Lyapunov exponent transversal to the synchronization manifold for $N=8$ Izhikevich neurons with electrical coupling under  {cluster} synchronization.}}
    \label{msfizk3}
\end{figure}

To check if all transverse modes are damped, we calculated the maximum Lyapunov exponent of Eq. \eqref{msfelecpartial},  which is given in Fig. \ref{msfizk3} as a function of $g_e$.

\begin{figure}[H]
    \centering
    \includegraphics[width=0.5\textwidth]{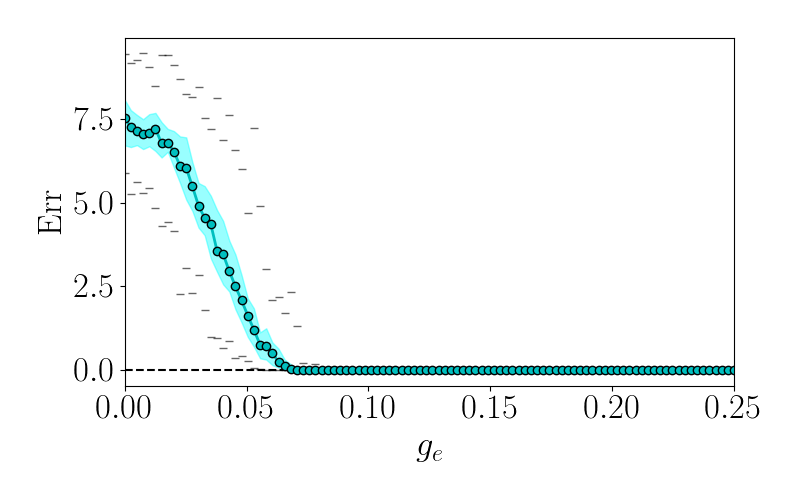}
    \caption{Synchronization error for Izhikevich model with electrical coupling,  {cluster} synchronization. The grey lines represent maximum and minimum errors, the shaded regions represent the first and third quartiles, and the circles the average over $100$ simulations.}
    \label{err_izk3cls}
\end{figure}

Fig. \ref{err_izk3cls} shows the synchronization error of the cluster synchronization state as a function of $g_e$. The error is computed as $\langle |x^i(t) - x_{m}^{i} (t)|\rangle$, where we take an average both in time and over the neurons $i$ in a cluster $m$. $x_{m}^{i}$ is the average state for the neurons in the cluster to which node $i$ belongs.

Fig. \ref{msfizk3} and Fig. \ref{err_izk3cls} show excellent agreement between the stability analysis and the synchronization error: the MLE associated with the transverse modes goes negative around $g_e \approx 0.073$, and the synchronization error vanishes around the same value.  

\subsection{Chemical coupling}

We now extend the linear stability study of cluster synchronized states to networks with chemical coupling, which is mediated by the adjacency matrix of the network. Consider a network with  $N$ nodes and adjacency matrix $A$, given a CSS with $M$ clusters encoded by an indicator matrix $Z$, the coarse-grained dynamics can be written as follows
\begin{align}
\dot{\mathbf{x}} &= (Z \otimes \mathbf{I}_q)\dot{\mathbf{s}},\label{clsync2} \quad\text{where} \\
\dot{\mathbf{s}} &= \mathbf{F}(\mathbf{s}) + g_e\mathbf{T}(\mathbf{s}) (A^{\pi} \otimes \mathbf{C})(\mathbf{s}) \ .
\end{align}

If the adjacency matrix of the quotient network is given by $A^{\pi} = Z^{+}AZ$, its diagonal entries correspond to self-loops in the quotient network whenever we have connections between nodes in the same cluster, and the partition encoded by $Z$ is an External Partition (EP). For EPs the number of connections from nodes in a cluster $C_i$ to nodes in a cluster $C_j$ depends only on $i,j$ \cite{schaub}.

The variational equation of Eq.(\ref{clsync2}) is

\begin{widetext}
\begin{equation}
    {\delta \dot{\mathbf{x}}} = \big\{ \sum_{m=1}^{M} [E^{m} \otimes ({D \mathbf{F}\big|_{\mathbf{s}^m}} - g_c P R^{m})] - [\sum_{m=1}^M (E^{m}A \otimes \mathbf{T}(\mathbf{s}^m))(\sum_{m=1}^M E^{m} \otimes {D\mathbf{K}\big|_{\mathbf{s}^m}})] \big\} \delta\mathbf{x} \ .
    \label{varchem}
\end{equation}
And after block-diagonalizing Eq. \eqref{varchem}, we obtain 
\begin{equation}
    \dot{\eta} = \big\{ \sum_{m=1}^{M} Q^{m} \otimes [{D \mathbf{F}\big|_{\mathbf{s}^m}} -  g_c \mathbf{G}R^m ] - ( Q^m\Gamma^A  \otimes \mathbf{T}(\mathbf{s}^m)(\sum_{m=1}^M Q^{m} \otimes {D\mathbf{K}\big|_{\mathbf{s}^m}})\big\} \eta \ .
\end{equation}    
\end{widetext}
   
where we defined

\begin{equation}    
    \mathbf{G}  =
    \begin{bmatrix}
    1 & 0    \\
    0 & 0  
    \end{bmatrix}
\quad\text{and}\quad
R^{m} = \sum_{n=1}^{M} A^{\pi}_{mn} \zeta(x_n) \ .
\end{equation}




{With these equation in hand,} we now consider the network of the previous subsection, {depicted in Fig. \ref{oclery}}, and the same indicator matrix $Z$ (Eq. \eqref{z1}), with chemical coupling between neurons. In this case, the adjacency matrix is given by
\begin{equation}
    A = 
    \begin{bmatrix}
    0 &   1 &   0 &   1 &   0 &   0 &   0 &   0  \\ 
    1 &   0 &   0 &   1 &   0 &   0 &   0 &   0  \\  
    0 &   0 &   0 &   1 &   0 &   0 &   0 &   0  \\
    1 &   1 &   1 &   0 &   1 &   1 &   0 &   0  \\ 
    0 &   0 &   0 &   1 &   0 &   0 &   1 &   1  \\ 
    0 &   0 &   0 &   1 &   0 &   0 &   1 &   1  \\ 
    0 &   0 &   0 &   0 &   1 &   1 &   0 &   1  \\ 
    0 &   0 &   0 &   0 &   1 &   1 &   1 &   0 \\
    \end{bmatrix} \ .
\end{equation}

\begin{figure}[H]
    \centering
    \includegraphics[width=0.5\textwidth]{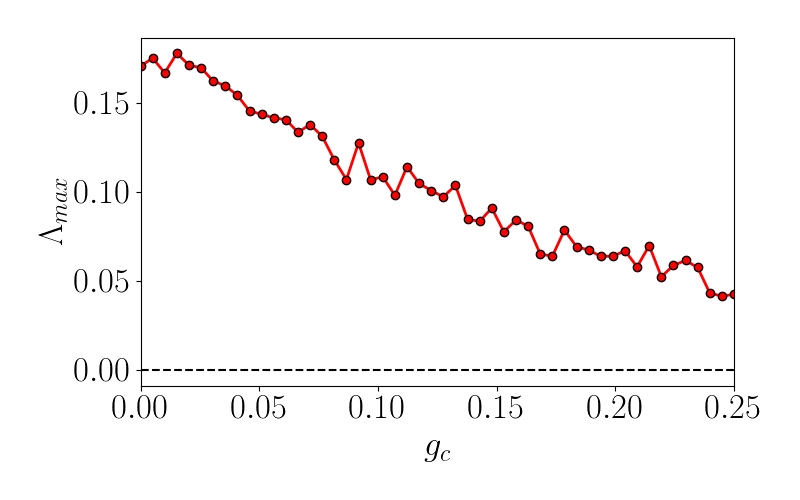}
    \caption{ {Maximum Lyapunov exponent transversal to the synchronization manifold for $N=8$ Izhikevich neurons with chemical coupling,  {cluster} synchronization.}}
    \label{msfb2}
\end{figure}

\begin{figure}[H]
    \centering
    \includegraphics[width=0.5\textwidth]{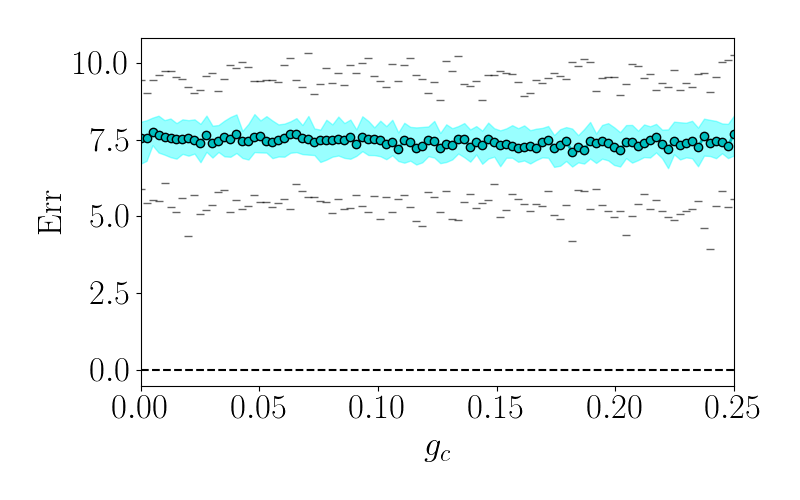}
    \caption{Synchronization error for Izhikevich model with chemical coupling,  {cluster} synchronization. The grey lines represent maximum and minimum errors, the shaded regions represent
the first and third quartiles, and the circles the average over $100$ simulations.}
    \label{err_izkoc2}
\end{figure}

Unlike the case with electrical coupling, here the CSS is not stable, as seen in Fig. (\ref{msfb2}), where the maximum Lyapunov exponent is positive in the range of $g_c$ considered. This analysis is supported by the synchronization error of this case (Fig. \ref{err_izkoc2}), which is greater than $0$ in the range of $g_c$ studied.

\subsection{Electrical and chemical coupling}

\begin{figure}[H]
    \centering
    \includegraphics[width=0.5\textwidth]{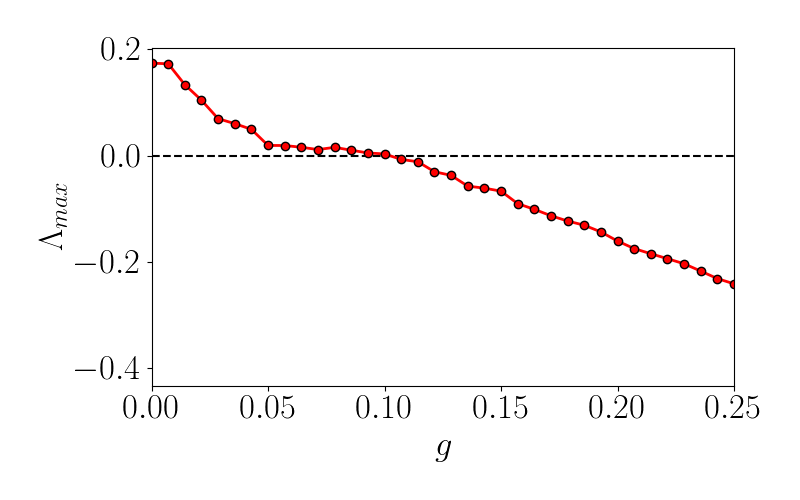}
    \caption{ {Maximum Lyapunov exponent transversal to the synchronization manifold for $N=8$ Izhikevich neurons with chemical and electric coupling,  {cluster} synchronization.}}
    \label{msf_izkoc3}
\end{figure}

\begin{figure}[H]
    \centering
    \includegraphics[width=0.5\textwidth]{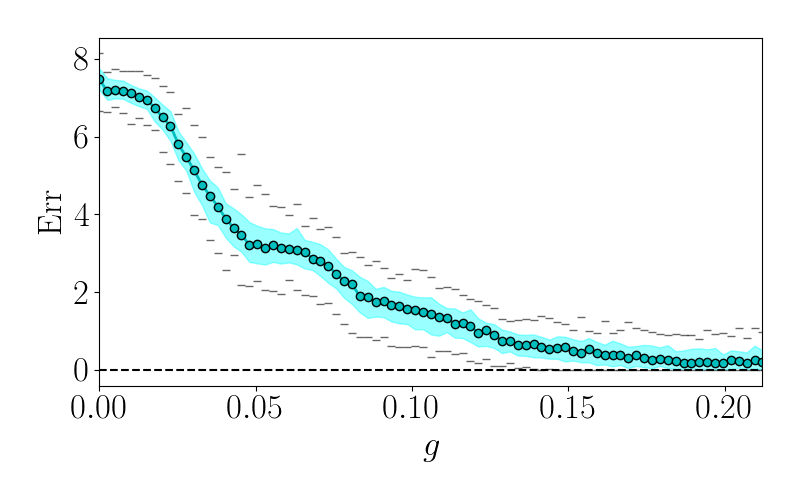}
    \caption{Synchronization error as a function of $g$ for Izhikevich model with electrical and chemical coupling. The grey lines represent maximum and minimum errors, the shaded regions represent
the first and third quartiles, and the circles the average over $100$ simulations.}
    \label{oc3_errbars}
\end{figure}

In the case where the network admits both electrical and chemical coupling, the dynamics of the CSS is governed by

\begin{align}
\dot{\mathbf{x}} = &(Z \otimes \mathbf{I}_q)\dot{\mathbf{s}},\label{clsync3} \quad\text{where} \\
\dot{\mathbf{s}} = \mathbf{F}(\mathbf{s}) - g_c\mathbf{T}(\mathbf{s}) &(A^{\pi} \otimes \mathbf{C})(\mathbf{s}) - g_e(L^{\pi} \otimes \mathbf{I}_q)\mathbf{G}(\mathbf{s}) \ .
\end{align}
where $Z$ is the indicator matrix of the CSS state, $A^{\pi}$ and $L^{\pi}$ are the quotient adjacency and Laplacian matrices, and for the sake of simplicity we take $g_c = g_e = g$. Putting together the results of the later subsections, the master stability function of a given CSS state can be written as
\begin{widetext}
\begin{equation}
 \hspace*{-1.4cm}\dot{\eta} = \big\{ \sum_{m=1}^{M} Q^{m} \otimes [{D \mathbf{F}\big|_{\mathbf{x}^m}} -  g_c \mathbf{G}R^m ] - g_c(\sum_{m=1}^M Q^m\Gamma^A  \otimes \mathbf{T}(\mathbf{x}^m)(\sum_{m=1}^M Q^{m} \otimes {D\mathbf{K}\big|_{\mathbf{x}^m}}) - g_e \sum_{m=1}^M \Gamma^L Q^m\otimes {D\mathbf{H}\big|_{\mathbf{x}^m}}\big\} \eta \ .
\label{msf_oc3}
\end{equation}
\end{widetext}
We proceed with the analysis of the network in Fig. \ref{oclery}, with indicator matrix \eqref{z1}. The MLE of Eq. \eqref{msf_oc3} is shown in Fig. \ref{msf_izkoc3}, where we notice that the introduction of electrical coupling in the network seems to induce the CSS, since the MLE becomes negative for  {$g > 0.102$}.
{The synchronization error is shown in Fig. \ref{oc3_errbars},}
the dotted line represents the {median}
over $100$ simulations and the shaded area is the {limit between the first and third quartiles, and the grey lines represent the upper and lower limits of the errors.} Although the lower bound of the
synchronization error goes to zero around $g \approx 0.125$, we
see that on average the system does not reach the synchronized solution. {While the initial conditions are similar to the ones used in Fig. \ref{err_izk3cls}, the introduction of chemical coupling results in a distinct outcome concerning the agreement between the MSF and synchronization error.} This sensibility to perturbations is characteristic of multistable systems \cite{ulrk}, and as discussed in \cite{huang}, a riddled basin of attraction can be the cause for such behavior, which in turn may be a consequence of the discontinuity of the differential equation. A thorough understanding of this phenomenon needs intense research which is outside the scope of this work.

\begin{figure}[H]
\includegraphics[width=.48\linewidth]{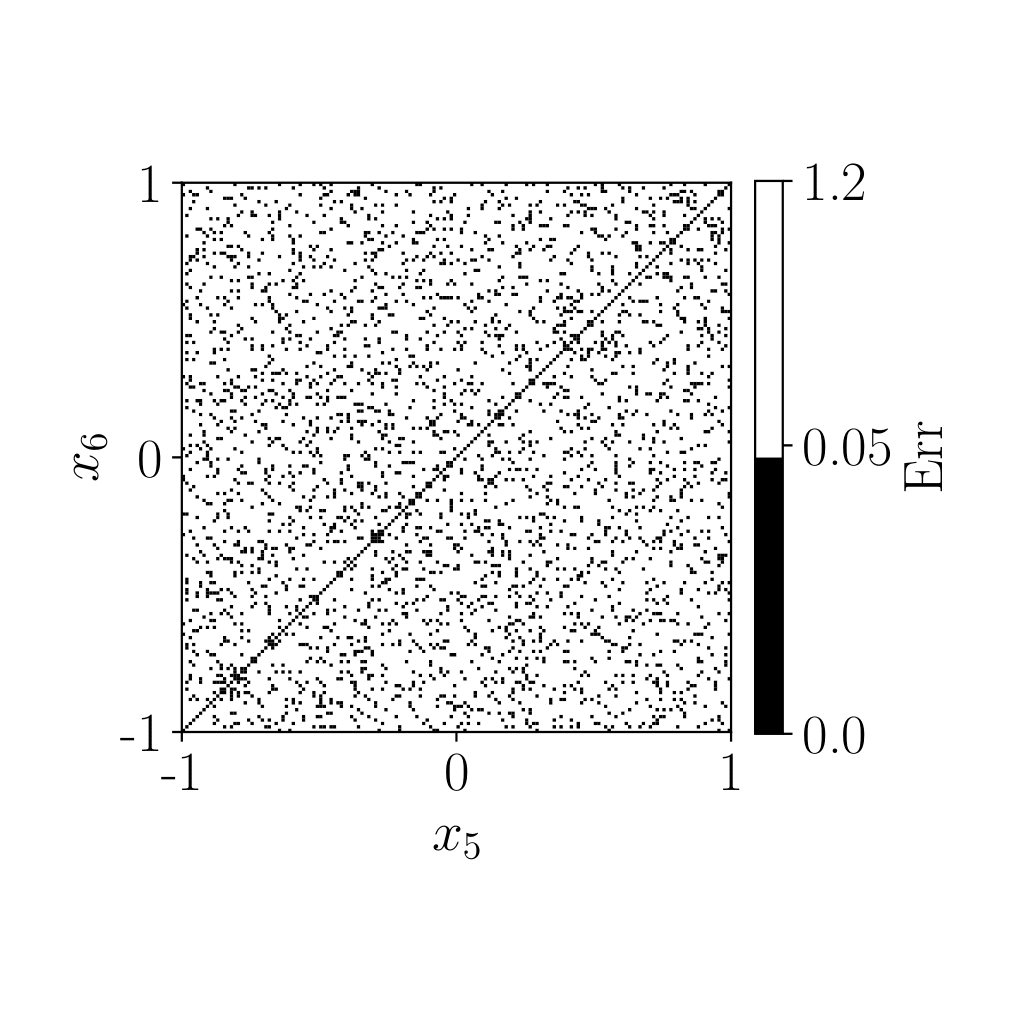}\hfill
\includegraphics[width=.48\linewidth]{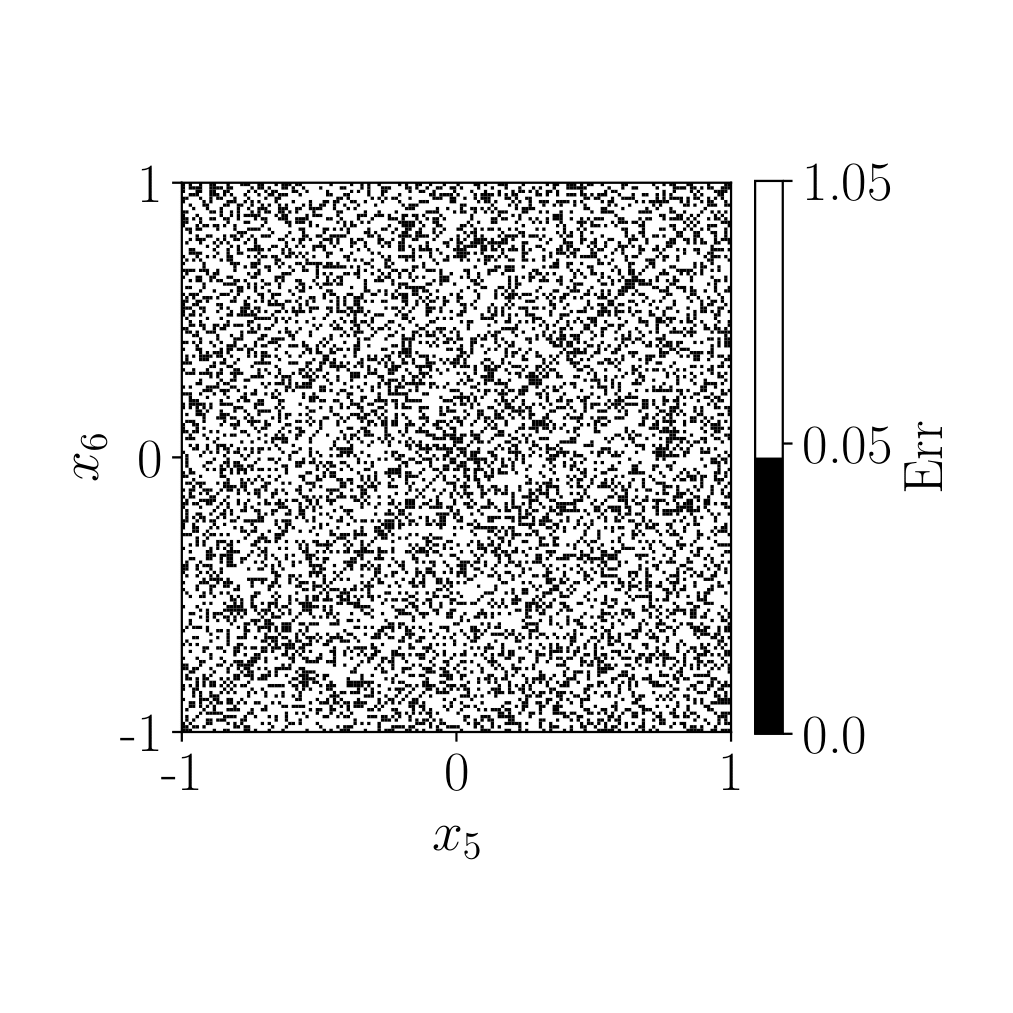}
    \caption{ {Riddled basin of attraction in $(x_5,x_6)$ plane for $g = 0.155$. (a) is calculated with $10.24\times10^4$ initial conditions in $x_5, x_6 \in [1,-1]$ and $y_{(5,6)} = -101.5$. The other nodes are initiate at $\mathbf{x}_{(1,2)} = [-59,-111]^T$,
$\mathbf{x}_3 = [-60,-110]^T$,
$\mathbf{x}_4 = [-58,-108]^T$ $\mathbf{x}_{(7,8)}= [-58,-109]^T$.
 (b) same as (a) for $g = 0.195$. We differentiate synchronized
solutions from non-synchronized solutions by setting a threshold at Err $= 0.05$, which are
represented by white and black dots respectively.}}
    \label{riddbasin}
\end{figure}

To verify that, we consider adding a disturbance in only one direction transverse to the synchronized solution, which means perturbing only one cluster. For example, if we take the solution
\begin{equation}\mathbf{x} = 
    \begin{bmatrix}
    \mathbf{s}_1\\
    \mathbf{s}_1\\
    \mathbf{s}_3\\
    \mathbf{s}_4\\
    \mathbf{s}_5\\
    \mathbf{s}_5\\
    \mathbf{s}_7\\
    \mathbf{s}_7
    \end{bmatrix}\ ,
    \quad\text{and add}\quad
    \mathbf{p}_{\perp} = \alpha
    \begin{bmatrix}
    0\\
    0\\
    0\\
    0\\
    1\\
    -1\\
    0\\
    0
    \end{bmatrix}\ \otimes
    \begin{bmatrix}
    1 \\
    1
    \end{bmatrix}, 
    \label{ptoc3}
\end{equation}

where $\alpha$ is a constant used to control the magnitude of the perturbation $\mathbf{p}_{\perp}$, we are pulling the first cluster out of synchronicity. 
Figure \ref{riddbasin} (a) shows how the synchronization error is sensible to initial conditions  {for $g = 0.155$. For simplicity, we differentiate synchronized solutions from non-synchronized solutions by setting a threshold at Err$= 0.05$, which are represented by white and black dots respectively. To calculate Err}, we consider initial conditions slightly perturbed as in Eq. \eqref{ptoc3}. The initial conditions for $\mathbf{x}_{(5,6)} = [x^{(5,6)},y^{(5,6)}]^T$ vary from $[1,1]$ to $[-1,-1]$, which is equivalent to $\alpha$ varying from $1$ to $-1$ in Eq. \eqref{ptoc3}. Along the diagonal $\mathbf{x}_5 = \mathbf{x}_6$ the system is always synchronized. In the off-diagonal region, we have both synchronized and non-synchronized solutions. {As in Fig. \ref{riddbasin0}, we see that} the basin is riddled with these two kinds of solutions. {The same calculation is depicted in Fig. \ref{riddbasin} (b), illustrating the riddle basin for $g = 0.195$. As anticipated from Fig. \ref{oc3_errbars}, the increased number of black dots signifies a higher proportion of points ending up in the synchronized basin.}

\begin{figure}[H]
   \centering
   \includegraphics[width=0.95\textwidth]{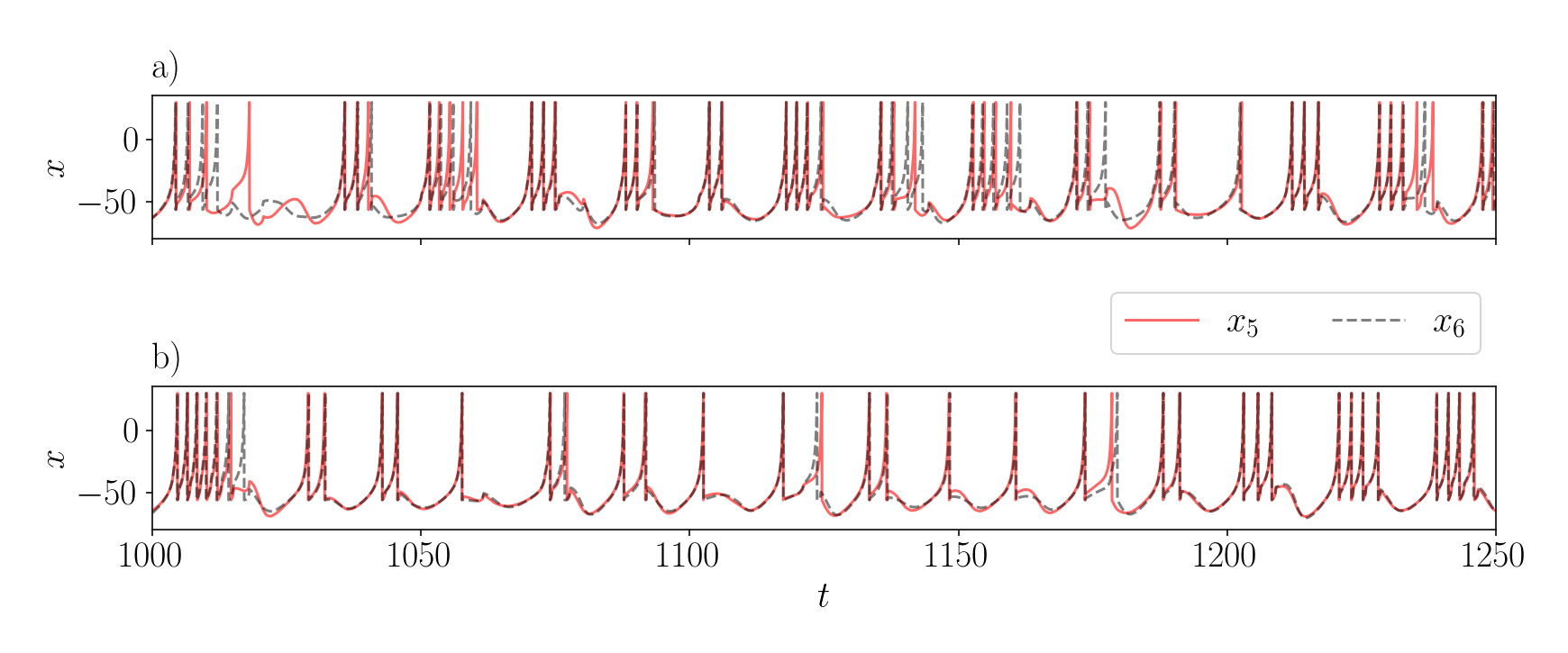}
   \caption{Time series of the $x$-variable from nodes $5$ (solid red) and $6$ (dashed black) for different initial conditions, both with $g=0.155$. (a) Incomplete synchronization Err $\approx 1.91$, (b) Err $\approx 0.50$.}
   \label{ts_oc3}
\end{figure}

{The solutions found in Fig. \ref{riddbasin} are shown in Fig. \ref{ts_oc3}. The solution in panel (a) yields a synchronization error greater than the one in panel (b). We notice that in both cases, there are finite windows of time where the system goes off synchronization, however, the solution in panel (a) seems to display more and longer windows. This type of intermittence is similar to the one discussed in \cite{gauthier1996}, and it is only found in the perturbed cluster formed by nodes 5 and 6, the remaining clusters stay synchronized throughout the simulation.}

\begin{figure}[H]
    \centering
    \includegraphics[width=0.5\textwidth]{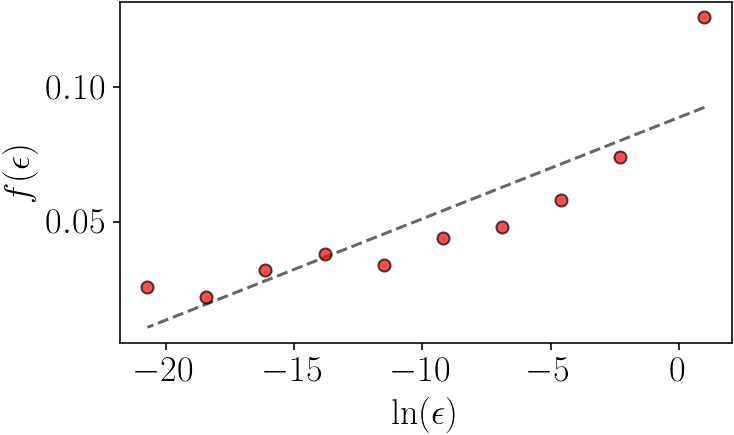}
    \caption{{The fraction of pairs of initial conditions that converge to different  {asymptotic solutions}, $f$, as a function of the distance between initial condition  $\varepsilon$ - Eq, \eqref{uncertain}. For $g = 0.1555$, the slope of the line that best fits the points yields an uncertainty exponent $\alpha = 0.007$.}}
    \label{fss_oc3}
\end{figure}

{To confirm the basin riddling, we compute the uncertainty exponent $\alpha$. Again, we take a pair of initial conditions  $\mathbf{x},\mathbf{x}'$ such that the distance between the two is $\epsilon = |\mathbf{x} - \mathbf{x}'|$ and verify if the initial conditions are uncertain or not. For each distance $\epsilon$, we simulate $1000$ pairs of initial conditions and save the uncertain fraction $f(\epsilon)$. The result is plotted in Fig. \ref{fss_oc3}, where the fitted line indicates an uncertainty exponent equal to $\alpha \approx 0.007$, and thus the basin dimension is equal to $d \approx 15.993$.}

\section{Conclusions}

We have calculated the MSF of networks of Izhikevich neurons, considering global and cluster synchronized states with electrical and chemical coupling schemes. To do so, we combined the MSF formalism with the use of saltation matrices, which allow the calculation of Lyapunov exponents of systems with discontinuities such as the Izhikevich model. For the networks studied, we found {that the synchronization error exhibits a nice agreement with the MSF analysis when only one type of coupling is considered. Moreover,} the addition of electrical coupling to a network with only chemical coupling induces synchronization for both global and cluster-synchronized states. However, the presence of both coupling schemes yielded a riddled basin of attraction. Due to this, even in the range where the MSF is negative and with nearly identical initial conditions concerning the CSS, the system can end up in a non-synchronized state.

 {This mismatch between the results expected by the MSF and the real output of the network
reinforces that, having a negative MLE for the synchronized state is \textit{necessary} but \textit{not sufficient}.}

{It will be interesting, for future work, to study not only networks with more elements but also the stability of synchronized solutions if we consider non-identical neurons and the presence of noise in the network. It would also be appealing to investigate why the riddled basins displayed by these systems appear as the MLE approaches zero from above, and not from below as it is usually reported in the literature.}

\section{Acknowledgements}
 {We would like to acknowledge Nicolas Rubido for revising the text, as well as express our gratitude for the valuable discussions held with Federico Bizzarri, Bruno Boaretto, and Eleonora Catsigeras.} R.P.A. acknowledges financing from Coordenação de Aperfeiçoamento de Pessoal
de Nível Superior – Brasil (CAPES), Finance Code 001. H.A.C. thanks ICTP-SAIFR and FAPESP grants 2021/14335-0 and
2021/11754-2 for partial support.\\

\appendix

\section{Derivation of MSF for the chemical coupling}\label{Appx}

Starting from Eq. \ref{chem} we define the $2\times2$ matrices 
\begin{equation}
    \mathbf{H} = 
    \begin{bmatrix}
    h  & 0 \\
    0  & 0
    \end{bmatrix}
\quad\text{and}\quad
    \mathbf{C}= 
    \begin{bmatrix}
    \zeta  & 0 \\
    0  & 0
    \end{bmatrix} \ ,
\end{equation}  
where $h$ and $\zeta$ are nonlinear operators
\begin{align}
    hx &= h(x) = x - v_s \ , \\
    \zeta x = \zeta(x) &=  [1 + \exp(-\epsilon(x-\theta))]^{-1} \ ,
\end{align}
and an operator $\mathbf{Dg}$, that returns a diagonal matrix whose elements are the components of the vector on which it acts. For example, for  $\mathbf{x}$:
\begin{equation}
    \mathbf{Dg} (\mathbf{x})  = 
    \mathbf{Dg} 
    \begin{bmatrix}
    x^1\\
    y^1 \\
    \vdots \\
    x^N  \\
    y^N 
    \end{bmatrix}
    =
    \begin{bmatrix}
    x^1  & 0 & ... & 0 & 0 \\
    0  & y^1 & ... & 0 & 0 \\
    \vdots  & \vdots & \ddots & \vdots & \vdots \\
    0  & 0 & ... & x^N & 0 \\
    0  & 0 & ... & 0 & y^N 
    \end{bmatrix} \ ,
\end{equation}
where in this notation $\mathbf{Dg}(\mathbf{x})$ reads $\mathbf{Dg}$ acts on $\mathbf{x}$.
We can write the vector encoding the synaptic coupling of a network of $N$ nodes as
\begin{equation}
    \mathbf{Dg}[(\mathbf{I}_N \otimes \mathbf{H})(\mathbf{x})](A \otimes \mathbf{C})(\mathbf{x}) \ .
\end{equation}

Moreover, it is convenient to introduce the matrices

\begin{equation}
    \mathbf{T} = \mathbf{Dg}(\mathbf{I}_N \otimes \mathbf{H})
\end{equation}
\begin{equation}
    \mathbf{K} = \mathbf{Dg}(\mathbf{I}_N \otimes \mathbf{C}) \ .
\end{equation}

With this notation, the dynamics of a network coupled through chemical synapses reads

\begin{equation}
    \mathbf{\dot{x}} = \mathbf{F}(\mathbf{x}) - g_c \mathbf{T}\mathbf{x}[(A \otimes \mathbf{C})(\mathbf{x})] \ .
\end{equation}
For example, if $N=1$,
\begin{equation*}
    \mathbf{T} \mathbf{x} = \mathbf{Dg}\Bigg[\Bigg(\mathbf{I}_1 \otimes 
    \begin{bmatrix}
    h & 0 \\
    0 & 0
    \end{bmatrix} \Bigg)
    \begin{bmatrix}
    x^1 \\
    y^1
    \end{bmatrix}\Bigg]
= \mathbf{Dg}
    \begin{bmatrix}
    h(x^1) \\
    0
    \end{bmatrix}
    \end{equation*}
    \begin{equation}
        = 
        \begin{bmatrix}
        x^1 - v_s & 0 \\
        0 & 0
        \end{bmatrix}
    \end{equation}
If we consider a network with $N=2$ neurons and an adjacency matrix given by
\begin{equation}
    A = 
    \begin{bmatrix}
        0 & 1 \\
        1 & 0 
    \end{bmatrix} \ ,
\end{equation}
the coupling term reads
\begin{align}           
\mathbf{T}\mathbf{x} &= \mathbf{Dg}\Bigg(\begin{bmatrix}
    1 & 0 \\
    0 & 1 
    \end{bmatrix}\otimes
    \begin{bmatrix}
    h & 0 \\
    0 & 0 
    \end{bmatrix}\Bigg)
    \begin{bmatrix}
    x^1  \\
    y^1  \\
    x^2  \\
    y^2  \\
    \end{bmatrix}
= \mathbf{Dg}           
    \begin{bmatrix}
    h & 0  & 0 & 0  \\
    0 & 0  & 0 & 0 \\
    0 & 0 & h & 0 \\
    0 & 0 & 0 & 0  \\
    \end{bmatrix}
    \begin{bmatrix}
    x^1  \\
    y^1  \\
    x^2  \\
    y^2  \\
    \end{bmatrix}
 \notag \\ 
\mathbf{Tx} &=   \mathbf{Dg}                    
    \begin{bmatrix}
    h(x^1)  \\
    0 \\
    h(x^2)  \\
    0  \\
    \end{bmatrix}
    =
    \begin{bmatrix}
    h(x^1) & 0 & 0 & 0\\
    0 & 0  & 0 & 0\\
    0 & 0 & h(x^2) & 0  \\
    0 & 0 & 0 & 0\\
    \end{bmatrix}
\end{align}
times
    \begin{equation}
    (A \otimes \mathbf{C})(\mathbf{x}) =
    \begin{bmatrix}
    0 & 0  & \zeta & 0  \\
    0 & 0  & 0 & 0 \\
    \zeta & 0  & 0 & 0 \\
    0 & 0  & 0 & 0  \\
    \end{bmatrix}
    \begin{bmatrix}
    x^1  \\
    y^1  \\
    x^2 \\
    y^2   \\
    \end{bmatrix}
    =
    \begin{bmatrix}
    \zeta(x^2)  \\
    0  \\
    \zeta(x^1) \\
    0   \\
    \end{bmatrix} 
\end{equation}
yielding
\begin{align}
\mathbf{T}(\mathbf{x})[(A \otimes \mathbf{C})(\mathbf{x})] &= 
    \begin{bmatrix}
    h(x^1) & 0 & 0 & 0\\
    0 & 0  & 0 & 0\\
    0 & 0 & h(x^2) & 0  \\
    0 & 0 & 0 & 0\\
    \end{bmatrix}
    \begin{bmatrix}
    \zeta(x^2)  \\
    0  \\
    \zeta(x^1) \\
    0   \\
    \end{bmatrix} \\
    \mathbf{T}(\mathbf{x})[(A \otimes \mathbf{C})(\mathbf{x})] &=
\begin{bmatrix}
    h(x^1)\zeta(x^2)  \\
    0  \\
    h(x^2)\zeta(x^1) \\
    0   \\
    \end{bmatrix} \ . 
\end{align} 

 Following the Pecora-Carroll analysis, the master stability function can be obtained via the variational equation for $\delta\mathbf{x}^i = \mathbf{x}^i - \mathbf{x}^s$. So first, we derive the synchronized solution $\mathbf{x}^s$, where $\mathbf{x}^i=\mathbf{x}^j$, $\forall$ $i,j$. If we take an adjacency matrix equal to Eq. \ref{izk_N4}, we have
\begin{equation}
        \mathbf{\dot{x}}^s = \mathbf{F}(\mathbf{x}^s) - g_c \mathbf{T}(\mathbf{x}^s)[(A \otimes \mathbf{C})(\mathbf{x}^s)] \ ,
        \label{appchem0} 
\end{equation}
the second term reads
\begin{equation}
\mathbf{Dg}\Bigg\{ \Bigg(\begin{bmatrix}
    1 & 0 & 0 & 0 \\
    0 & 1 & 0 & 0 \\
    0 & 0 & 1 & 0 \\
    0 & 0 & 0 & 1
    \end{bmatrix}\otimes
    \begin{bmatrix}
    h & 0 \\
    0 & 0 
    \end{bmatrix}\Bigg)
    \begin{bmatrix}
    x^s  \\
    y^s  \\
    \vdots \\
    x^s  \\
    y^s  \\
    \end{bmatrix}
    \Bigg\}
    \Bigg\{
        \Bigg(
    \begin{bmatrix}
       0 & 1 & 0 & 1 \\
       1 & 0 & 1 & 0 \\
       0 & 1 & 0 & 1 \\
       1 & 0 & 1 & 0
    \end{bmatrix}
    \otimes
    \begin{bmatrix}
    \zeta & 0\\
    0     & 0
    \end{bmatrix}
    \Bigg)
    \begin{bmatrix}
    x^s \\
    y^s \\
    \vdots \\
    x^s\\
    y^s
    \end{bmatrix}
    \Bigg\}
\end{equation} 
which, after application of $\mathbf{Dg}$ becomes:
\begin{equation}
g_c
    \begin{bmatrix}
       h(x^s) & 0 & 0 & 0 & 0 & 0 & 0 & 0 \\
       0 & 0 & 0 & 0 & 0 & 0 & 0 & 0 \\
       0 & 0 & h(x^s) & 0 & 0 & 0 & 0 & 0 \\
       0 & 0 & 0 & 0 & 0 & 0 & 0 & 0 \\
       0 & 0 & 0 & 0 & h(x^s) & 0 & 0 & 0 \\
       0 & 0 & 0 & 0 & 0 & 0 & 0 & 0 \\
       0 & 0 & 0 & 0 & 0 & 0 & h(x^s) & 0 \\
       0 & 0 & 0 & 0 & 0 & 0 & 0 & 0 \\
    \end{bmatrix}
    \Bigg(
    \begin{bmatrix}
       0 & 1 & 0 & 1 \\
       1 & 0 & 1 & 0 \\
       0 & 1 & 0 & 1 \\
       1 & 0 & 1 & 0
    \end{bmatrix}
    \otimes
    \begin{bmatrix}
    \zeta & 0\\
    0     & 0
    \end{bmatrix}
    \begin{bmatrix}
    x^s \\
    y^s \\
    \vdots \\
    x^s\\
    y^s
    \end{bmatrix}
    \Bigg)
\end{equation}
simplifying we obtain:
\begin{equation}
g_c
    \begin{bmatrix}
       h(x^s) & 0 & 0 & 0 & 0 & 0 & 0 & 0 \\
       0 & 0 & 0 & 0 & 0 & 0 & 0 & 0 \\
       0 & 0 & h(x^s) & 0 & 0 & 0 & 0 & 0 \\
       0 & 0 & 0 & 0 & 0 & 0 & 0 & 0 \\
       0 & 0 & 0 & 0 & h(x^s) & 0 & 0 & 0 \\
       0 & 0 & 0 & 0 & 0 & 0 & 0 & 0 \\
       0 & 0 & 0 & 0 & 0 & 0 & h(x^s) & 0 \\
       0 & 0 & 0 & 0 & 0 & 0 & 0 & 0 \\
    \end{bmatrix}
    \begin{bmatrix}
       \zeta(x^s) + \zeta(x^s)  \\
       0  \\
       \zeta(x^s) + \zeta(x^s)  \\
       0  \\
       \zeta(x^s) + \zeta(x^s)  \\
       0  \\
       \zeta(x^s) + \zeta(x^s)  \\
       0  \\
    \end{bmatrix}    
    \end{equation}
which is symply
\begin{equation}
    g_c k_n \mathbf{T}(\mathbf{x}^s)\mathbf{C}(\mathbf{x}^s)    
    \label{appchem1}
\end{equation}
where in this case, the number of links that each node has is $k_n = 2$.
Now, we linearize $\mathbf{x}^i$ around the synchronized state $\mathbf{x}^s$

\begin{equation}
    \mathbf{x}^i \approx \mathbf{F}(\mathbf{x}^s) + D\mathbf{F}\big|_{\mathbf{x}^s}\delta\mathbf{x}^i - g_c D \Big[ \mathbf{T}(\mathbf{x}^i) \sum_{j} a_{ij}\mathbf{C}(\mathbf{x}^j)\Big] \delta\mathbf{x}^k \ ,
\end{equation}
where $k = (i,j)$ depending on whether the Jacobian matrix $D$ acts on a function of $\mathbf{x}^i$ or $\mathbf{x}^j$. Evaluating the last term yields

\begin{equation}
    D [\mathbf{T}(\mathbf{x}^i) \sum_j a_{ij}\mathbf{C}(\mathbf{x}^j)] \delta\mathbf{x}^k = 
    D
    \begin{bmatrix}
    h(x^i) \sum_j a_{ij}\zeta(x^j) \\
    0
    \end{bmatrix} \delta\mathbf{x}_k
\end{equation}
\begin{equation}
   =
    \begin{bmatrix}
    \frac{\partial h(x^i)}{\partial x^i} \sum_j a_{ij}\zeta(x^j) & 0 \\
    0   & 0
    \end{bmatrix}\delta\mathbf{x}^i
    +
    \begin{bmatrix}
    h(x^i) \sum_j a_{ij}\frac{\partial \zeta(x^j)}{\partial x^i} & 0 \\
    0   & 0
    \end{bmatrix}\delta\mathbf{x}^j
    \end{equation}
    \begin{equation}
   =
    \begin{bmatrix}
    \frac{\partial h(x^i)}{\partial x^i} k_n\zeta(x^j) & 0 \\
    0   & 0
    \end{bmatrix}\Bigg|_{\mathbf{x}^s}\delta\mathbf{x}^i
    +
    \begin{bmatrix}
    h(x^i) \sum_j a_{ij}\frac{\partial \zeta(x^j)}{\partial x^i} & 0 \\
    0   & 0
    \end{bmatrix}\Bigg|_{\mathbf{x}^s}\delta\mathbf{x}^j
    \end{equation}
which is
\begin{equation}
    g_c k_n D\mathbf{H}\big|_{\mathbf{x}^s} \mathbf{K}(\mathbf{x}^s)\delta\mathbf{x}^i - g_c \mathbf{T}(\mathbf{x}^s) \sum_j a_{ij}D\mathbf{C}\big|_{\mathbf{x}^s} \delta\mathbf{x}^j  \ . 
\end{equation}
Then, we can write $\delta\mathbf{x}^i$ as
\begin{align}
    \delta\mathbf{x}^i = D\mathbf{F}\big|_{\mathbf{x}^s} - g_c k_n D\mathbf{H}\big|_{\mathbf{x}^s} \mathbf{K}(\mathbf{x}^s)\delta\mathbf{x}^i  - g_c \mathbf{T}(\mathbf{x}^s) \sum_j a_{ij}D\mathbf{C}\big|_{\mathbf{x}^s} \delta\mathbf{x}^j  \ . 
\end{align}
and finally, we write $\delta\mathbf{x} = [\delta\mathbf{x}^1, ..., \delta\mathbf{x}^N]^T$ as
\begin{align}
    \delta\dot{\mathbf{x}} = \{[\mathbf{I}_N \otimes (D\mathbf{F}\big|_{\mathbf{x}^s} - g_ck_nD\mathbf{H}\big|_{\mathbf{x}^s} \mathbf{K}(\mathbf{x}^s))] - g_c A \otimes \mathbf{T}(\mathbf{x}^s)D\mathbf{C}\big|_{\mathbf{x}^s} \} \delta\dot{\mathbf{x}} \ .
\end{align} 

The diagonalization of $A$ leads to the desired block diagonalized variational equation:
\begin{align}
    \dot{\eta} = \{[\mathbf{I}_N \otimes (D\mathbf{F}\big|_{\mathbf{x}^s} - g_ck_nD\mathbf{H}\big|_{\mathbf{x}^s}\mathbf{K}(\mathbf{x}^s))]  - g_c \Gamma \otimes \mathbf{T}(\mathbf{x}^s)D\mathbf{C}\big|_{\mathbf{x}^s} \} \eta \ .
    \label{msfizk2ap}
\end{align}

\bibliography{refs}

\end{document}